\documentclass[oneside,12pt]{article}
\usepackage{eurosym}
\usepackage{amsmath}
\usepackage{amsfonts}
\usepackage{graphicx,epsfig}
\usepackage{epstopdf}
\usepackage{amssymb}

\begin{document}

\title{\textbf{\LARGE Statefinder diagnostic for modified Chaplygin gas
cosmology in $f(R,T)$ gravity with particle creation}}
\author{{\large J. K. Singh}$^{1}${\large , Ritika Nagpal}$^{2}$, {\large S.
K. J. Pacif}$^{3}$ \\
$^{1,2}$\textit{Department of Mathematics,}\\
\textit{\ \ Netaji Subhas Institute of Technology,}\\
\textit{\ \ Faculty of Technology, University of Delhi,}\\
\textit{\ \ New Delhi 110 078, India}\\
$^{3}$\textit{Centre for Theoretical Physics,}\\
\textit{\ Jamia Millia Islamia,\ }\\
\textit{New Delhi 110 025, India}\\
\textit{\ \ jainendrrakumar@rediffmail.com}$^{1}$ \textit{,
ritikanagpal.math@gmail.com}$^{2}$, \ \\
\ \ \textit{shibesh.math@gmail.com}$^{3}$\\
}
\maketitle

{\footnotesize \vskip0.4in \noindent \textbf{Abstract} In this paper, we
have studied flat Friedmann--Lema\^{\i}tre--Robertson--Walker (FLRW) model
with modified Chaplygin gas (MCG) having equation of state $p_{m}=A\rho -%
\frac{B}{\rho ^{\gamma }}$, where $0\leq A\leq 1$, $0\leq \gamma \leq 1$ and 
$B$ is any positive constant in }${\footnotesize f(R,T)}${\footnotesize \
gravity with particle creation. We have considered a simple parametrization
of the Hubble parameter $H$ in order to solve the field equations and
discussed the time evolution of different cosmological parameters for some
obtained models showing unique behavior of scale factor. We have also
discussed the statefinder diagnostic pair $\{r,s\}$ that characterizes the
evolution of obtained models and explore their stability. The physical
consequences of the models and their kinematic behaviors have also been
scrutinized here in some detail.}

{\footnotesize \vskip0.2in \noindent \textbf{Keywords} }${\footnotesize %
f(R,T)}${\footnotesize \ theory, Chaplygin gas, Particle creation,
Parametrization \newline
 }

\newpage

\section{\protect\Large Introduction}

\noindent \qquad The current understanding of our observed Universe reveal
that we live in an expanding Universe which is some $13.8$ billion years old
that have been originated with a bang from a phase of very high density and
temperature. Ever since this has been expanding. For a long time, it was
assumed that either the expansion is gradually slow down caused by inward
pull of gravity and would ultimately come to a halt after which Universe
start to contract into a big crunch or the Universe expands eternally.
However, it was at the end of twentieth century, the cosmological
observations \cite{RIE, PERM} of type Ia supernovae revealed that the
Universe might be expanding with an acceleration. The unexpected discovery
surprised the cosmologists because the idea of cosmic acceleration was
against the standard predictions of decelerating expansion caused by
gravity. Later on it is predicted that the three quarters of the volume of
Universe consists of some exotic stuff termed as \textquotedblleft Dark
energy\textquotedblright\ (DE) with highly negative pressure causing the
acceleration. The subsequent discoveries in this direction gave more and
more evidences for for a flat, dark energy dominated accelerating Universe.
However, there are alternative way to explain the acceleration e.g. to
modify the theory of gravity. In both the cases we confront new physics.

The Quantum field theory (QFT) and general theory of relativity (GTR)
suggested the most viable candidate of DE - cosmological constant $\Lambda $ 
\cite{PEE, SAH, PAD} introduced by Einstein. Though the model with a
cosmological constant known as $\Lambda $CDM model is well versed it has
some shortcomings \cite{Wein}. Dynamical models of dark energy was proposed
in the past few years with some effective candidates of DE \cite{quint1,
quint2, quint3, quint4, kessen1, kessen2, spint, tachy1, tachy2, quintom1,
quintom2, quintom3, quintom4, chamel1, chamel2} explaining some other
observational features of the Universe. For a brief reviews on various dark
energy models one can see \cite{Sami, Bamba}. One of the most prospective
candidate of DE is Chaplygin gas (CG) \cite{chapl}. In order to understand
the cosmic acceleration, Chaplygin gas is a simple characterization among
the various class of dark energy models. The EoS of CG cosmological model is
given by $p=-\frac{A}{\rho }$. Some of the inspiring and remarkable
attributes of CG is that it can discuss the dark sector of the Universe with
a single fluid component, leading to the unified models of DE and DM \cite%
{kame, fab, bento}. In the different eras of the Universe, CG plays a binary
role: as in the early phase of the evolution of the Universe it acts like a
dust matter and like a cosmological constant in the late time Universe \cite%
{bilic}. CG exhibits an easy distorting of the standard $\Lambda $CDM model.
One of the most prominent characteristic of CG cosmological model is that it
provides a desirable phase transition from decelerated cosmic expansion to
accelerated one. Some of the conceptual achievements of CG models has given
in \cite{gorini}. Further, CG has positive and bounded squared velocity of
sound which is not obvious for its negative pressure fluids. 
\begin{equation}
c_{s}^{2}=\frac{\partial p}{\partial \rho }=\frac{A}{\rho ^{2}}\text{.}
\label{1a}
\end{equation}%
\newline
\qquad Several generalization of CG has been proposed in the literature \cite%
{dev, sche, debu} due to its inspiring features. A generalized version of CG
(GCG) is specified by its EoS, $p=-\frac{A}{\rho ^{\gamma }}$, $0\leq \gamma
\leq 1$, in which both DE and DM are just two different sides of a single
exotic fluids. In accordance to have more consistency with observational
data GCG was further extended to Modified Chaplygin gas (MCG) \cite{bena,
sin1, sin2}. The MCG was presented with an EoS 
\begin{equation}
p_{m}=A\rho -\frac{B}{\rho ^{\gamma }},  \label{1}
\end{equation}%
where $0\leq A\leq 1$, $0\leq \gamma \leq 1$, and $B$ is any positive
constant. MCG EoS consists of two parts, the first part recovers an ordinary
perfect fluid with a linear barotropic EoS, and the second part connects
pressure to some power of the inverse of energy density. In EoS of MCG, the
case $B=0$ leads to standard perfect fluid while it reduces to the GCG EoS
with $A=0$.

Matter creation in the Universe is also an important concept to be worth
noting after the pioneering work of Parker \cite{parker1, parker2, parker3,
parker4, parker5}. The proposal was that it is the gravitational field
acting on quantum vacuum responsible for the continuous creation of
radiation and matter in an expanding Universe. Leonard Parker \cite{park}
suggested that the massless or massive particles production are not occurred
in radiation or matter dominate eras \cite{parker6, parker7, parker8, full}.
To understand the concept of this particle production there are two general
approaches. The first one is the technique of adiabatic vacuum state \cite%
{parker6, parker7, parker8, full} and the second one is the technique of
instantaneous Hamiltonian diagonalization \cite{pavl, grib1, grib2, grib3}.
The particle creation scenario has many other aspects including the future
deceleration phase in the Universe, existence of emergent Universe and
chance of phantom Universe without the inclusion of phantom field. Also, to
describe the current accelerating expansion of the Universe, a logical way
can be the particle creation mechanism that based on QFT without the
inclusion of any exotic component (DE) and the very concept was first
proposed by Prigogine \cite{prig, prigo}.

In this paper, we have constructed some FLRW models with modified Chaplygin
gas (MCG) EoS in $f(R,T)$ gravity with particle creation. Here, we have
considered the simple parametrization of the Hubble parameter $H$ proposed
in \cite{SKJP}, to obtain some deterministic solutions to Einstein field
equations (EFE). The physical behavior of energy density, matter pressure
and the pressure due to particle creation are discussed for three different
models. The paper is organized in seven sections as follows. Sect. 1 is of
introductory nature. In sect. 2, the modification of GTR \textit{i.e.}, $%
f(R,T)$ gravity is discussed. In sect. 3, we have studied the field
equations and solutions in which the energy density, matter pressure, and
the pressure due to particle creation for three different models so
obtained. In sect. 4, we have discussed a general probe for the expansion
dynamics of the Universe using the statefinder diagnostic pair $\{r,s\}$. In
sect. 5, we have discussed the stability criteria imposed on the velocity of
sound $C_{s}^{2}$. Some distances in cosmology have been analyzed through
kinematic tests in sect. 6 for all the models. Finally, the concluding
remarks for the obtained models have been discussed in sect. 7.

\section{${\protect\Large f(R,T)}${\protect\Large \ gravity}}

\noindent \qquad \qquad The $f(R,T)$ theory is the modification of the
general theory of relativity (GTR), where $R$ and $T$ are scalar curvature
and the trace of stress energy-momentum tensor respectively \cite{tfs}. The
total gravitational action in $f(R,T)$ gravity is of the form

\begin{equation}  \label{2}
S=\frac{1}{16\pi G}\int \sqrt{-g}[f(R,T)+L_{m}]d^{4}x,
\end{equation}%
where $L_{m}$ is the matter Lagrangian density and $g $ the metric
determinant. Taking the variation the action (\ref{1}) into account \textit{%
w.r.t.} the metric tensor components yields

\begin{equation}  \label{3}
G_{\mu\nu}+\left( g_{\mu\nu}\square -\nabla _{\mu}\nabla _{\nu}\right)
=[8\pi +2f^{\prime }(T)]T_{\mu\nu}+2[f^{\prime }(T)p+\frac{1}{2}%
f(T)]g_{\mu\nu},
\end{equation}

\noindent for which it is assumed that $f(R,T)=R+2f(T)$, where primes denote
differentiation \textit{w.r.t.} the argument. Now, we take $f(T)=\lambda T$
, with $\lambda $ a constant. This is the simplest non-trivial functional
form of the function $f(R,T)$, which includes non-minimal matter-geometry
coupling within $f(R,T)$ formalism. Moreover, it benefits from the fact that
GR is retrieved when $\lambda=0$. Here, we consider perfect fluid as the
matter source of the Universe and therefore, the energy-momentum tensor of
matter Lagrangian can be taken as 
\begin{equation}  \label{4}
T_{\mu\nu}=(\rho +p_{m})u_{\mu}u_{\nu}-p_{m}g_{\mu\nu},
\end{equation}%
where $\rho $ and $p_{m},$ are the dominant energy density and matter
pressure of the cosmic fluid, respectively. $u^{\mu}=(0,0,0,1)$ is the
components of the four velocity vector in the co-moving coordinate system
which satisfies the conditions $u^{\mu}u_{\mu}=1 $ and $u^{\mu}\nabla
_{\nu}u_{\mu}=0 $. We choose the perfect fluid matter as $L_{m}=-p_{m} $ in
the action (\ref{1}).

\section{Field equations and solutions}

\noindent \qquad \qquad The background metric satisfying the cosmological
principle considered here in the form of the flat FLRW metric 
\begin{equation}
ds^{2}=dt^{2}-a^{2}(t)\sum_{i=1}^{3}(dx_{i}^{2})\text{.}  \label{5}
\end{equation}%
\newline

We assume the matter content in the Universe filled with perfect fluid. In
the presence of particle creation, the energy-momentum tensor of the perfect
fluid (\ref{4}) takes the form 
\begin{equation}
T_{\mu \nu }=(\rho +p_{m}+p^{c})u_{\mu }u_{\nu }-(p_{m}+p^{c})g_{\mu \nu },
\label{6}
\end{equation}%
where $p^{c}$ is the pressure due to particle creation which depends on the
particle production rate. The trace of the stress-energy-momentum in the
influence of particle creation is 
\begin{equation}
T=\rho -3(p_{m}+p^{c})\text{.}  \label{7}
\end{equation}

The gravitational field equations in the above background are obtained as 
\begin{equation}
3H^{2}=8\pi \rho +f(T)+2(\rho +p_{m}+p^{c})f^{\prime }(T),  \label{8}
\end{equation}%
\begin{equation}
2\dot{H}+3H^{2}=-8\pi (p_{m}+p^{c})+f(T).  \label{9}
\end{equation}%
\newline

Equations (\ref{7}), (\ref{8}) and (\ref{9}) yield 
\begin{equation}
3H^{2}=\left( 8\pi +3\lambda \right) \rho -\lambda p_{m}-\lambda p^{c},
\label{10}
\end{equation}%
\begin{equation}
2\dot{H}+3H^{2}=\lambda \rho -\left( 8\pi +3\lambda \right) p_{m}-\left(
8\pi +3\lambda \right) p^{c}),  \label{11}
\end{equation}%
where an overhead dot indicates the derivative \textit{w.r.t.} cosmic time $%
t $. Here, $p^{c}$ is the particle creation pressure which is a dynamic
pressure that depends on production rate of particles. For, if $p^{c}$ is
negative, this may compel the accelerating expansion of the Universe. Due to
the firm constraints foist by local gravity measurements \cite{ellis,
peebles, hagiw}, the ordinary particle production is much limited and the
radiation component has practically no influence on the acceleration. Some
precise attention can be made to a process called \textit{adiabatic}
particle production which means, particles and also the entropy ($S$) are
produced in a space-time but the entropy per particle $(\sigma =\frac{S}{N})$
(or specific entropy) is remains constant. For this case, the creation
pressure reads \cite{adiab1, adiab2, adiab3} 
\begin{equation}
p^{c}=-\frac{(\rho +p_{m})\Gamma }{3nH}\text{.}  \label{12}
\end{equation}%
\qquad

To fulfil our goal, here in this paper, we use the parameterization of $%
\Gamma $ \cite{nunes, pan, chak, saha, chakr, dutt, chakra, haro, deharo} as 
$\Gamma =3nH\eta $, which a source term indicating the production $(\Gamma
>0)$ of particles and annihilation $(\Gamma <0)$ of the particles, $n$
refers to the particle number density and $H$ is the Hubble parameter (HP).
The constant $\eta $ $\in \lbrack 0,1]$. The term $n\eta >0$ can be denoted
as a free parameter of the model that characterizes the particle production
process. For, if $n\eta =0$ then, it means that there is no matter creation
and for if high $n\eta $ then, there is high production of particle. But, in
all of these cases, $\Gamma /3H\leq 1$. By use of $\Gamma =3nH\eta $ in (\ref%
{12}), the particle creation pressure $p^{c}$ takes the form 
\begin{equation}
p^{c}=-(\rho +p_{m})\eta \text{.}  \label{13}
\end{equation}

Using equation (\ref{13}) in equation (\ref{10}) and (\ref{11}), we obtain 
\begin{equation}
3H^{2}=\left[ 8\pi +\left( 3+\eta \right) \lambda \right] \rho +\lambda
\left( \eta -1\right) p_{m},  \label{14}
\end{equation}%
\begin{equation}
2\dot{H}+3H^{2}=\left[ 8\pi \eta +\left( 1+3\eta \right) \lambda \right]
\rho +\left[ 8\pi \left( \eta -1\right) +3\lambda \left( \eta -1\right) %
\right] p_{m}\text{.}  \label{15}
\end{equation}

Now, we have two field equations containing three variables $\rho ,$ $p_{m}$%
, and $a$ in terms of $H$ and its derivative. We need to specify the matter
content in the Universe which can be classified by its pressure. In the
introduction, we have briefly discussed the importance of MCG in describing
the late-time acceleration of the Universe having EoS \textbf{(\ref{1})} for
which the field equations (\ref{14}) and (\ref{15}) reduce to 
\begin{equation}
3H^{2}=\left[ 8\pi +\left( 3+\eta \right) \lambda +A\left( \eta -1\right)
\lambda \right] \rho -B\left( \eta -1\right) \lambda \rho ^{-\gamma },
\label{16}
\end{equation}%
\begin{eqnarray}
2\dot{H}+3H^{2} &=&\left[ 8\pi \eta +\left( 1+3\eta \right) \lambda +8\pi
A\left( \eta -1\right) +3A\left( \eta -1\right) \lambda \right] \rho  \notag
\label{17} \\
&&-\left[ 8\pi \left( \eta -1\right) B+3\left( \eta -1\right) \lambda B%
\right] \rho ^{-\gamma }.
\end{eqnarray}%
Eliminating $H^{2}$ from equation (\ref{16}) and (\ref{17}), we can write
the field equations in a single evolution equation as 
\begin{equation}
\dot{H}=\left( \eta -1\right) \left( 1+A\right) \left( 4\pi +\lambda \right)
\rho -\left( \eta -1\right) \left( 4\pi +\lambda \right) B\rho ^{-\gamma },
\label{18}
\end{equation}%
which can alternatively be written as a polynomial equation in terms of $%
\rho $ given by 
\begin{equation}
\rho ^{\gamma +1}+\left[ \frac{\dot{H}}{\left( 1-\eta \right) \left(
1+A\right) \left( 4\pi +\lambda \right) }\right] \rho ^{\gamma }-\frac{B}{%
\left( 1+A\right) }=0\text{.}  \label{19}
\end{equation}

Equation (\ref{19}) can be solved for $\rho $ for a known scale factor $a(t)$
by providing particular values of $\gamma $. In literature, one can find
number of parametrization of scale factor $a(t)$ and its higher order
derivative terms \textit{i.e.}, first order derivative - Hubble parameter $%
H(t)$ and second order derivative - deceleration parameter $q(t)$. For a
recent review on various parametrization, see \cite{SKJP}. The technique
commonly known as the `model independent way' to study dark energy models.
Although the arbitrary constrain on any cosmological parameter seems to be
an adhoc choice, this do not violate the background theory anyway for which
the `model independent way' or the `parametrization' is a fiducial technique
to study the models with an extra degree of freedom \textit{i.e.}, dark
energy. Clearly, the dynamical behaviors of model depend on the functional
form of assumed parameter. Following the same technique one can consider
some specific parametrization of scale factor or its higher order
derivatives for a comparative study. In this present work, we assume the
simple and convenient form of Hubble parameter considered in \cite{SKJP} as%
\begin{equation}
H(t)=\frac{\beta t^{k_{1}}}{\left( t^{k_{2}}+\alpha \right) ^{k_{3}}},
\label{20}
\end{equation}%
where $\alpha ,$ $\beta \neq 0,$ $k_{1},$ $k_{2},$ $k_{3}$ are real
constants. $\alpha $, $\beta $ both may have the dimensions of time that can
reduce to many known models under one umbrella by specifying $k_{1},$ $%
k_{2}, $ $k_{3}$. However, we won't consider all the models that have been
discussed in \cite{SKJP}. Here, we consider only three models showing
completely different evolution of scale factor \textit{e.g.}, the power law
model \cite{PLC1} (that also leads to Berman's model of constant
deceleration parameter \cite{BERMAN1}), linearly varying deceleration
parameter (LVDP) model \cite{AKARSU} and a non singular model leading to a
time varying deceleration parameter \cite{q-ASSRP} for a comparative study
with MCG equation of state and particle creation in $f(R,T)$ theory of
gravity.

\subsection{Model-I}

\qquad For $k_{1}=-1,$ $k_{3}=0,$ $\forall $ $k_{2}$ in equation (\ref{20}),
we obtain the power law model with $H(t)=\frac{\beta }{t}$ and $%
a(t)=Ct^{\beta }$, where $\beta >0$ is dimensionless model parameter. The
deceleration parameter $q=\frac{1}{\beta }-1\,$, which is constant
throughout the evolution. Equation (\ref{19}) reduces to

\begin{equation}  \label{21}
\rho ^{\gamma +1}+\left[ \frac{\beta }{\left( \eta -1\right) \left(
1+A\right) \left( 4\pi +\lambda \right) }\right] \frac{1}{t^{2}}\rho
^{\gamma }-\frac{B}{1+A}=0,
\end{equation}
which can not be solved for general $\gamma $. By providing some particular
values of constants $A $, $B $, $\lambda $, $\eta $ and $\beta $, we show
the evolution of energy density $\rho $ for different values of $\gamma $
which gives different evolution as can be seen from equation (\ref{1}).
Similarly, the evolution of pressure is shown in the plot below
corresponding to the functional forms of $\rho $.

\begin{figure}[tbph]
\begin{center}
$%
\begin{array}{c@{\hspace{.1in}}cc}
\includegraphics[width=2.2 in]{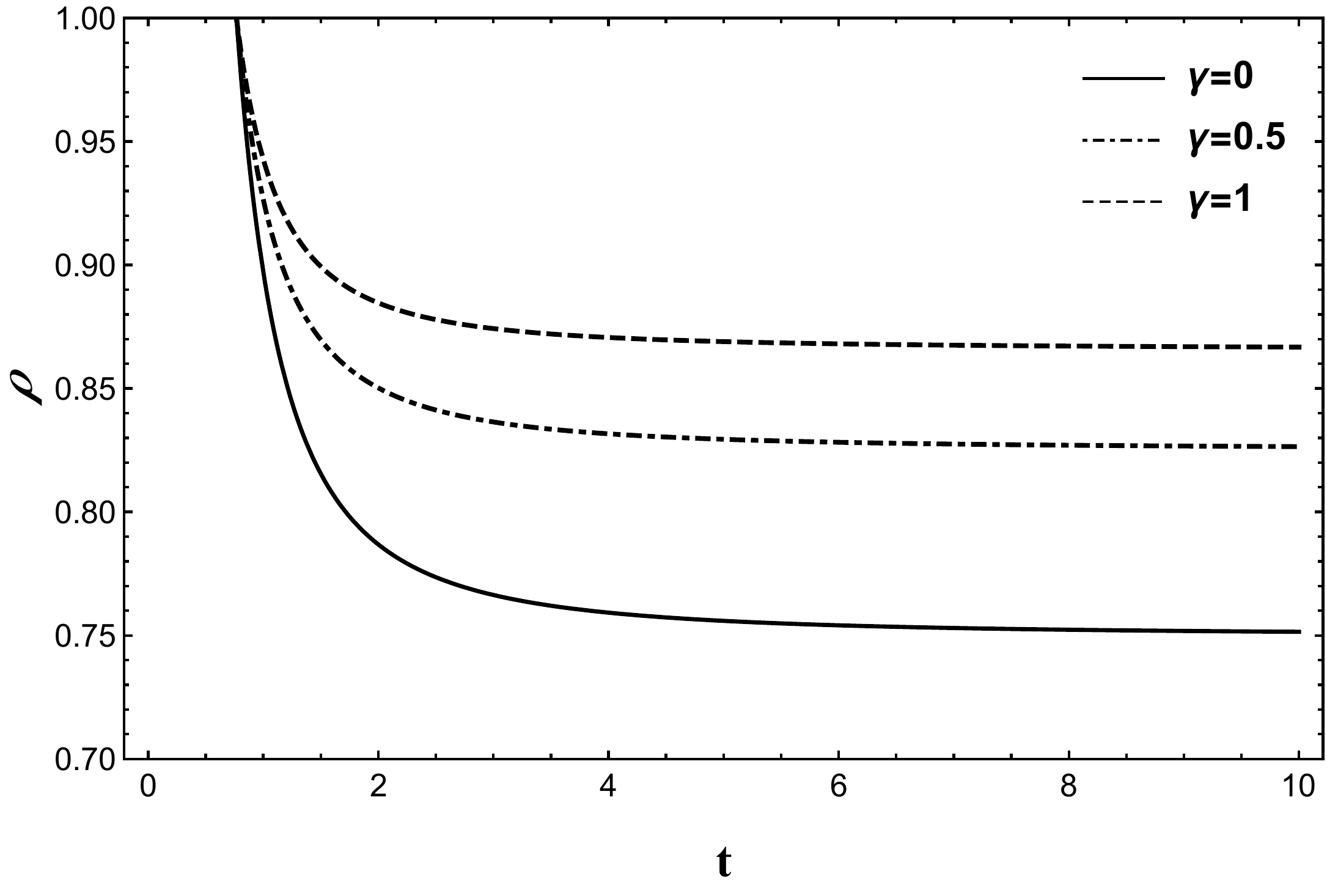} & 
\includegraphics[width=2.2
in]{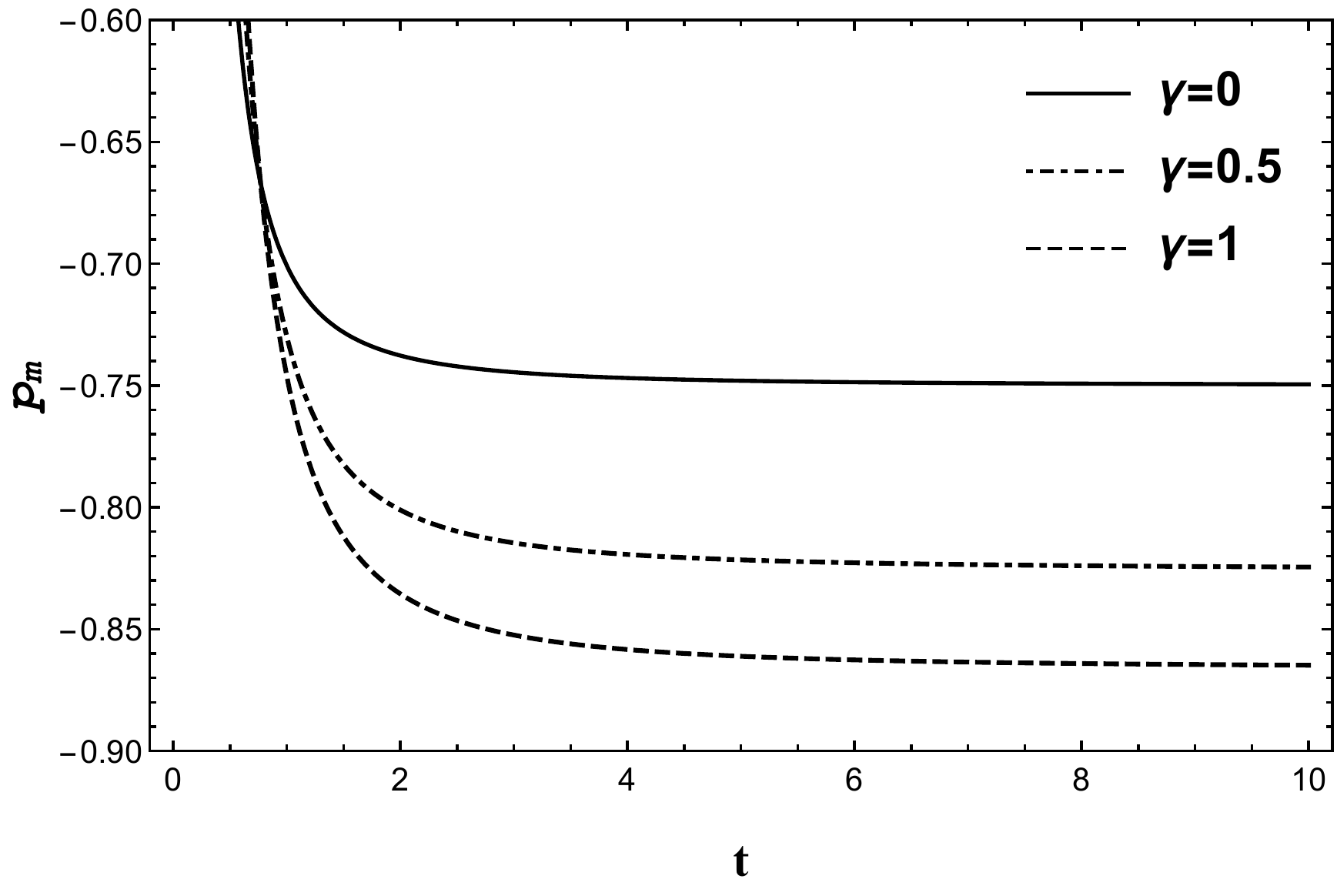} & \includegraphics[width=2.2 in]{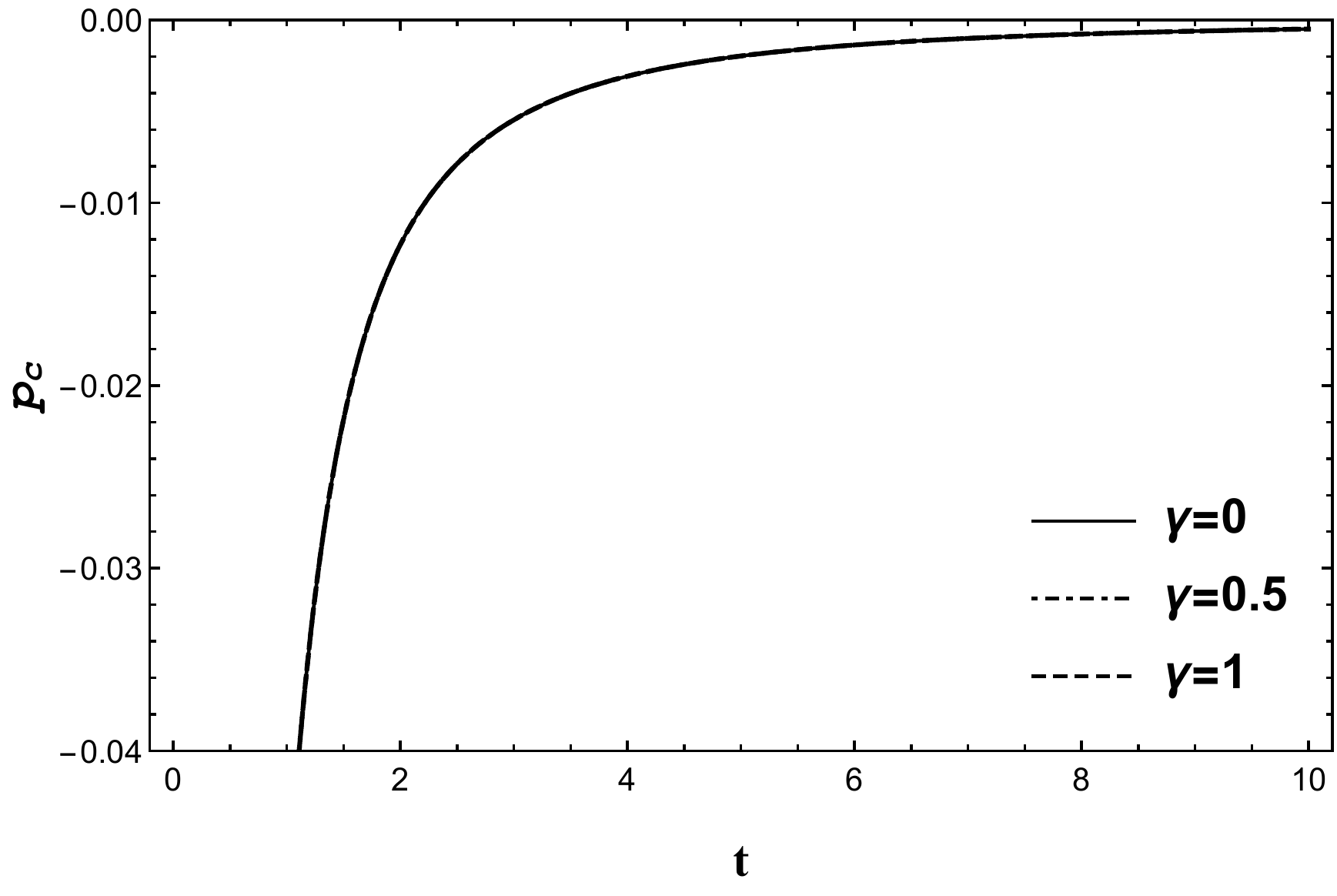} \\ 
\mbox (a) & \mbox (b) & \mbox (c)%
\end{array}%
$%
\end{center}
\caption{\textit{\ The plots of }$\protect\rho $\textit{, }$p_{m}$\textit{\
and }$p_{c}$\textit{\ Vs. time }$t$\textit{\ for the Model-I with }$\mathit{%
A=}\frac{1}{3}$\textit{, }$\mathit{B=1}$\textit{, }$\protect\lambda =1$%
\textit{, }$\protect\eta =0.25$\textit{\ and }$\protect\beta =2$\textit{\
for the particular values of }$\protect\gamma =0$, $0.5$, $1$\textit{.}}
\end{figure}

From Fig. 1a, we can observe that the energy density $\rho $ is very very
high ($\rightarrow \infty $) initially for all the three cases of $\gamma =0$%
, $0.5$, $1$. As the time unfolds, the energy density falls rapidly and
attain a constant value in the late time Universe. In Fig. 1b, we see that
the matter pressure decreases as time increases and remains negative which
represents the accelerated cosmic expansion for all the three different
values of $\gamma $. Fig. 1c depicts that the particle creation pressure $%
p_{c}$ is initially negative, overlying in all the three different values of 
$\gamma $ and tends to zero in the late-time. Here, we observe that the
universe is accelerated expanding since $p^{c} $ is always negative, and the
rate of particle production are initially high and later decreases to almost
negligible corresponding to constant $\rho $. \\[10pt]

\subsection{Model-II}

\qquad For $k_{1}=-1,$ $k_{2}=1,$ $k_{3}=1$ in equation (\ref{20}), we have $%
H(t)=\frac{\beta }{t\left( t+\alpha \right) }$ and $a(t)=C\left( \frac{t}{%
t+\alpha }\right) ^{\frac{\beta }{\alpha }}$, where $\alpha ,\beta <0$ are
model parameters, both have dimensions of time. The deceleration parameter $%
q(t)=-1+\frac{\alpha }{\beta }+\frac{2}{\beta }t$ vary linearly and shows
phase transition from deceleration to acceleration in the near past.
Equation (\ref{19}) takes the form here as

\begin{equation}
\rho ^{\gamma +1}+\left[ \frac{\beta }{\left( \eta -1\right) \left(
1+A\right) \left( 4\pi +\lambda \right) }\right] \frac{2t+\alpha }{t\left(
t+\alpha \right) ^{2}}\rho ^{\gamma }-\frac{B}{1+A}=0.  \label{22}
\end{equation}%
\newline

For some particular $\gamma $, equation (\ref{22}) can be solved by
providing suitable values to $A$, $B$, $\lambda \,$, $\eta $, $\alpha $ and $%
\beta $. The evolution of energy density $\rho $ and pressures are shown in
the following figures. 
\begin{figure}[tbph]
\begin{center}
$%
\begin{array}{c@{\hspace{.1in}}cc}
\includegraphics[width=2.2 in]{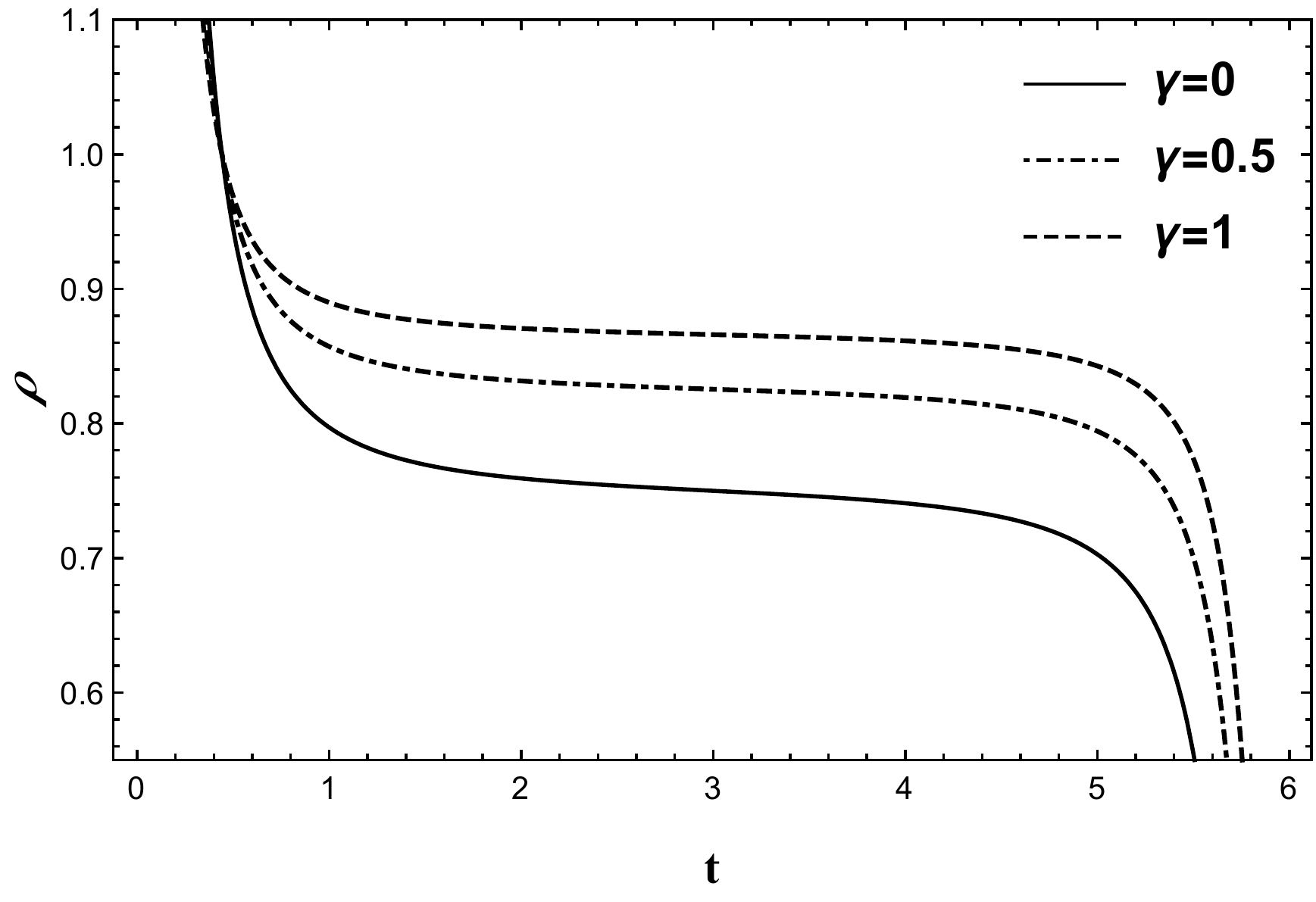} & 
\includegraphics[width=2.2
in]{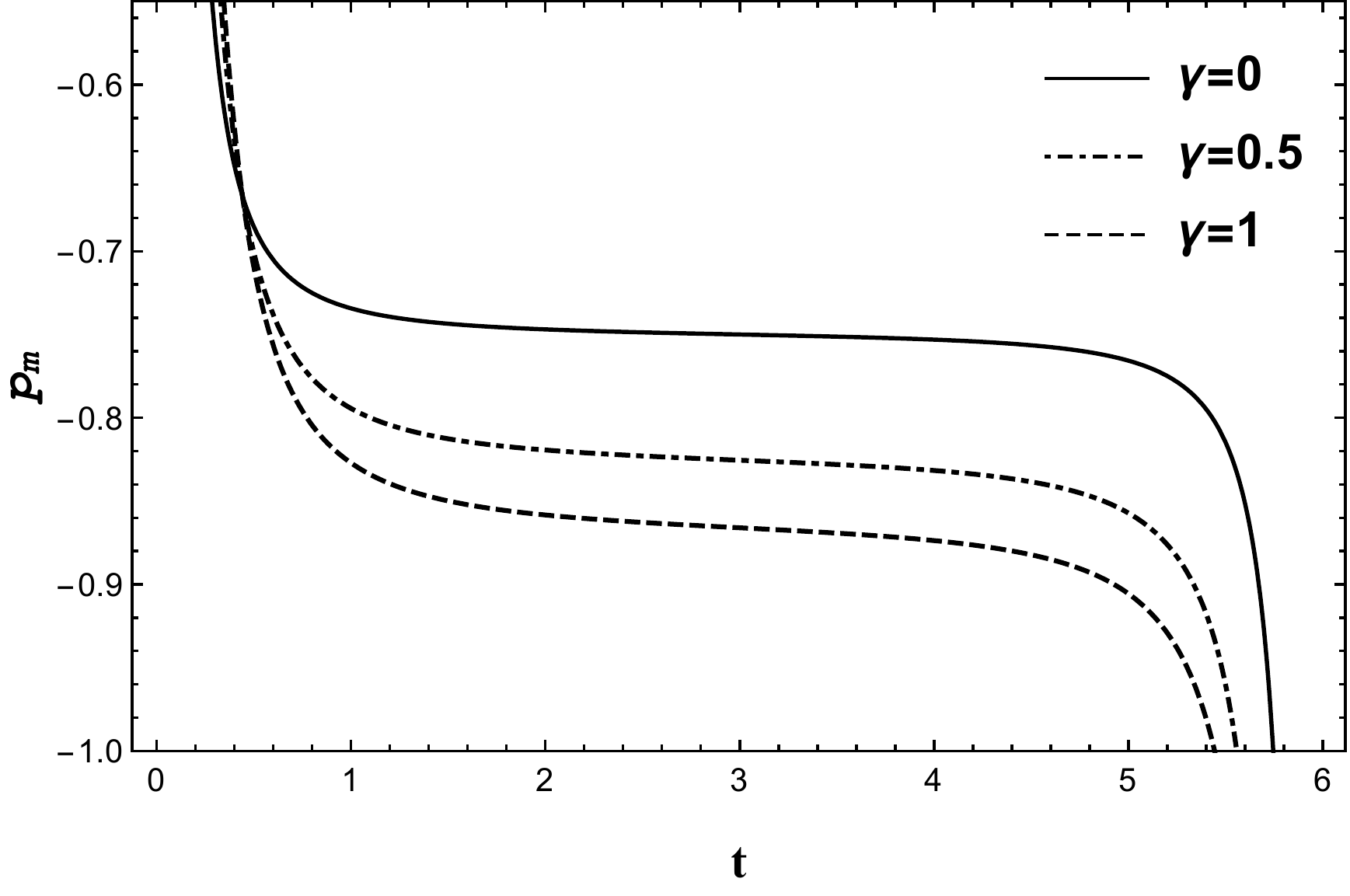} & \includegraphics[width=2.2 in]{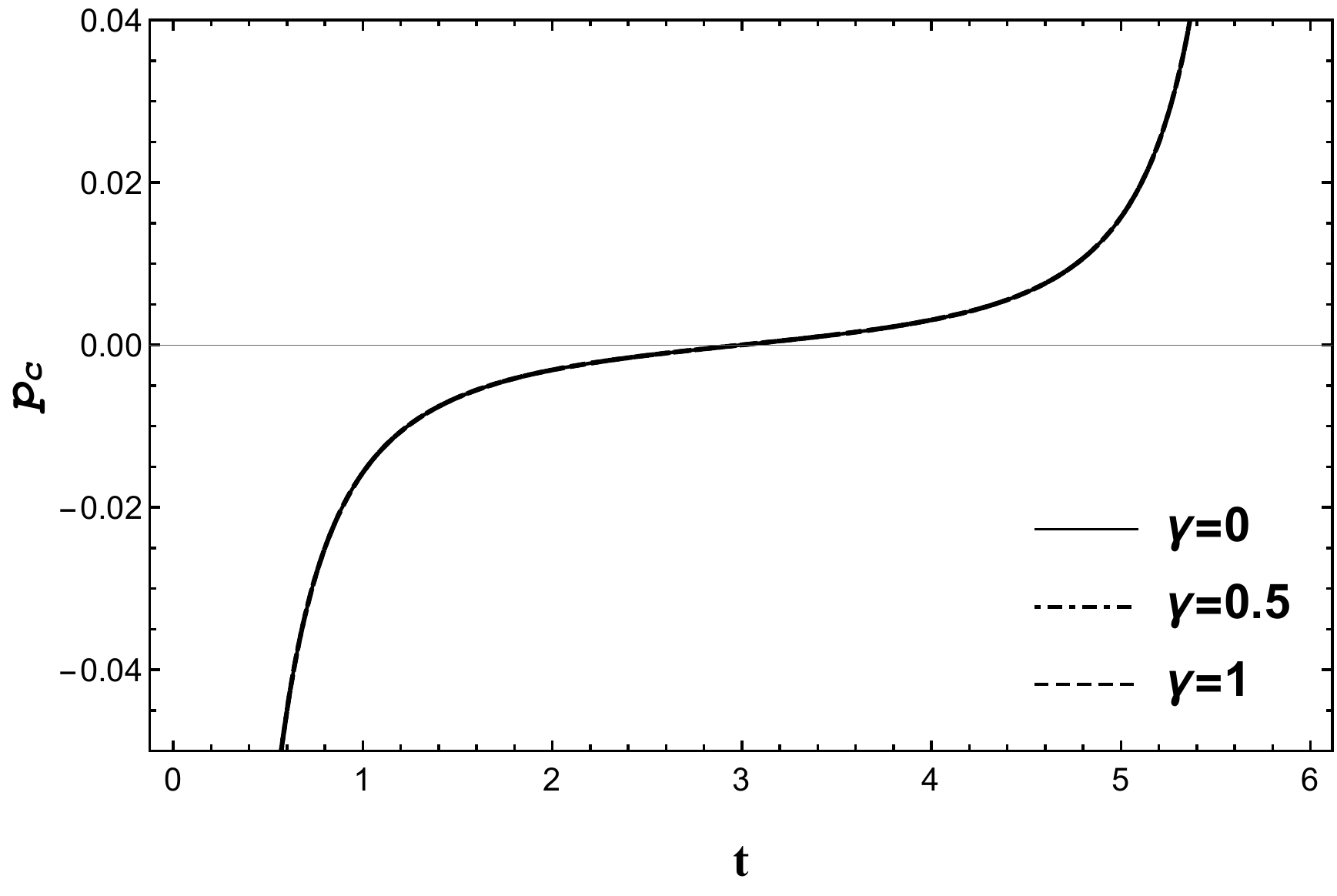} \\ 
\mbox (a) & \mbox (b) & \mbox (c)%
\end{array}%
$%
\end{center}
\caption{\textit{The plots of }$\protect\rho $\textit{, }$p_{m}$\textit{\
and }$p_{c}$\textit{\ Vs. time }$t$\textit{\ for the Model-II with }$\mathit{%
A=}\frac{1}{3}$\textit{, }$\mathit{B=1}$\textit{, }$\protect\lambda =1$%
\textit{, }$\protect\eta =0.25$\textit{\ and }$\protect\alpha =-6$ \& $%
\protect\beta =-4$\textit{\ for the particular values of }$\protect\gamma =0$%
, $0.5$, $1$\textit{.}}
\end{figure}
\vspace{0.2in}

In Fig. 2a, the plot of energy density $\rho $ starts with a very large
value and after a short period of time it decreases promptly, remains
constant for some time in the Universe, follows an immediate fall and
subsequently diverges towards negative in future representing a future
singularity at time $t=\alpha $. Fig. 2b depicts that the matter pressure $%
p_{m}$ recedes initially and gradually decreases with slow rate then after
some time falls down rapidly and remains negative throughout the evolution
of the Universe in all the three cases. In Fig. 2c, we observe that the
particle creation pressure $p_{c}$ overlap for all the three different
values of $\gamma $, initially negative and increases with time, tends to
zero then expands with increasing rate and remains positive in the late time
Universe. Here, we observe that the universe is accelerated expanding and
the rate of particle production are initially high and later decreases to
almost negligible, and again increases to attain high rate of particle
production at $t=6$. \newline

\subsection{Model-III}

\qquad For $k_{1}=1,$ $k_{2}=2,$ $k_{3}=1$ in equation (\ref{20}), we obtain
the non-singular bouncing model with $H(t)=\frac{\beta t}{t^{2}+\alpha }$
and $a(t)=C\left( t^{2}+\alpha \right) ^{\frac{\beta }{2}}$, where $\alpha
,\beta >0$ are model parameters. $\alpha $ has the dimension of square of
time and $\beta $ is dimensionless. The deceleration parameter in this case
comes out to be $q(t)=-1+\frac{1}{\beta }-\frac{\alpha }{\beta }\frac{1}{%
t^{2}}$. We have from equation (\ref{19}),

\begin{equation}
\rho ^{\gamma +1}+\left[ \frac{\beta }{\left( 1-\eta \right) \left(
1+A\right) \left( 4\pi +\lambda \right) }\right] \frac{\alpha -t^{2}}{\left(
t^{2}+\alpha \right) ^{2}}\rho ^{\gamma }-\frac{B}{1+A}=0.  \label{23}
\end{equation}%
\newline

The evolution of energy density $\rho $ and pressures are shown in the
following figures for different values of $\gamma $ and suitable choice of $%
A $, $B$, $\lambda \,$, $\eta $, $\alpha $ and $\beta $.

\begin{figure}[tbh]
\begin{center}
$%
\begin{array}{c@{\hspace{.1in}}cc}
\includegraphics[width=2.2 in]{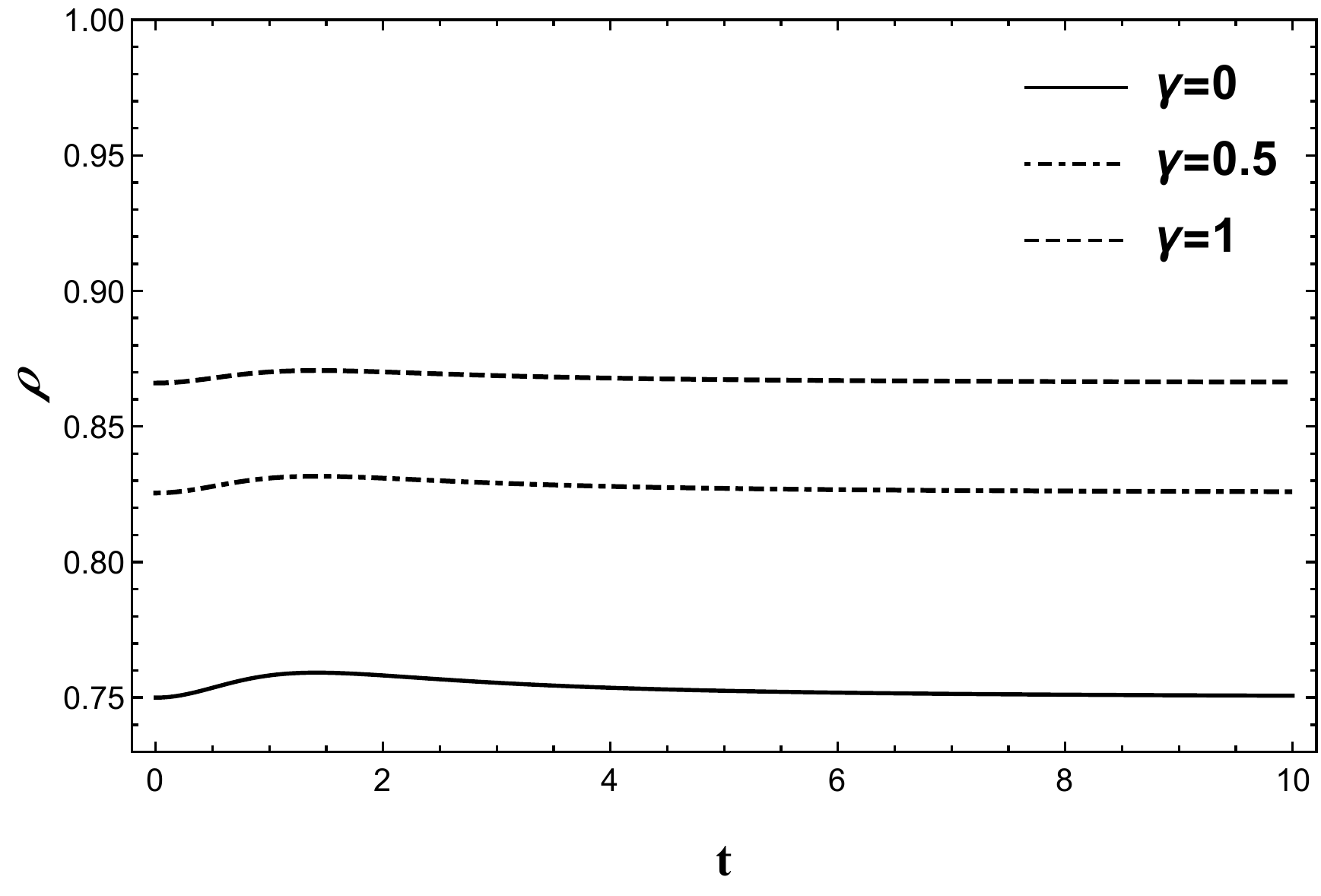} & 
\includegraphics[width=2.2
in]{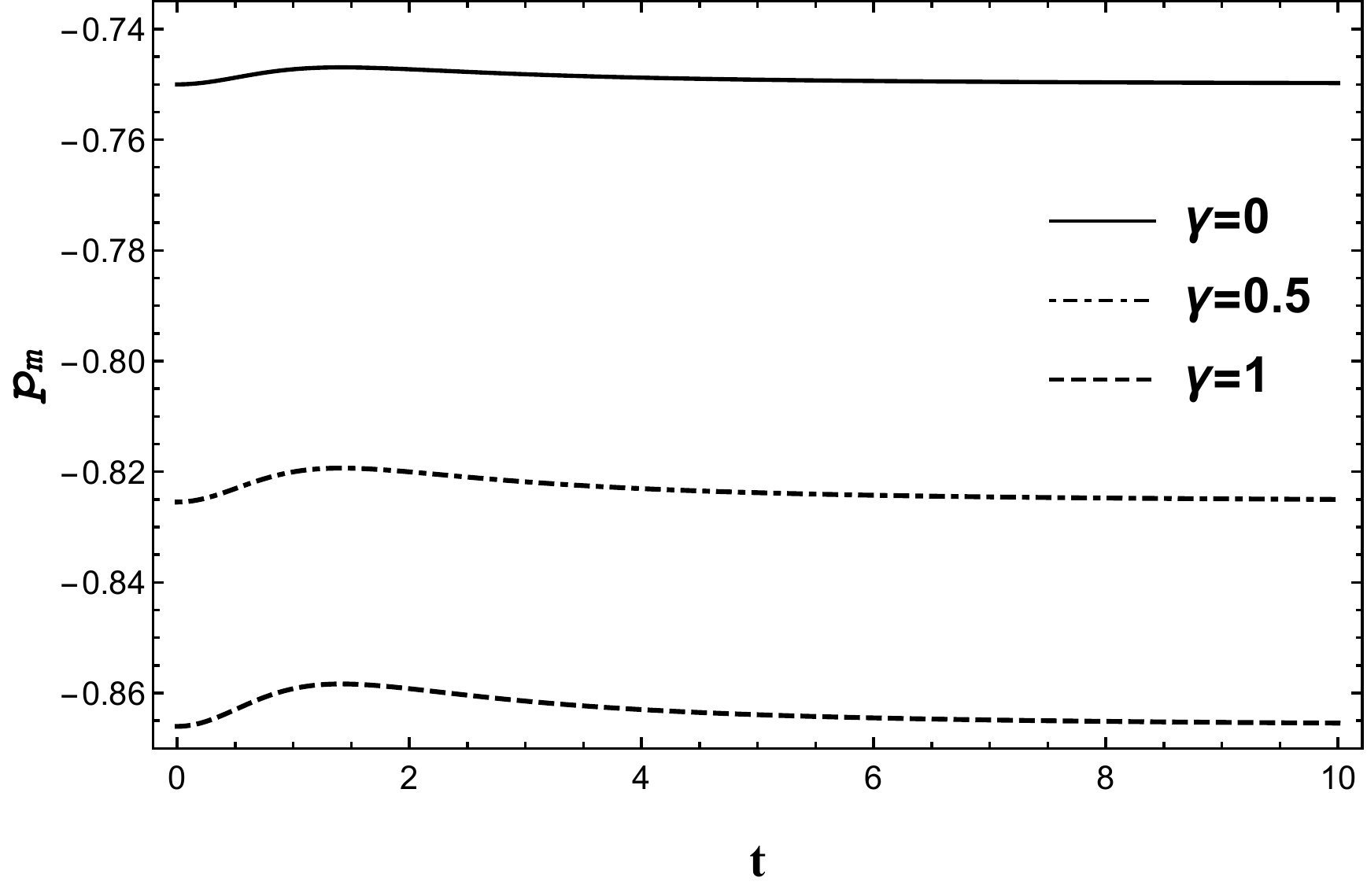} & \includegraphics[width=2.2 in]{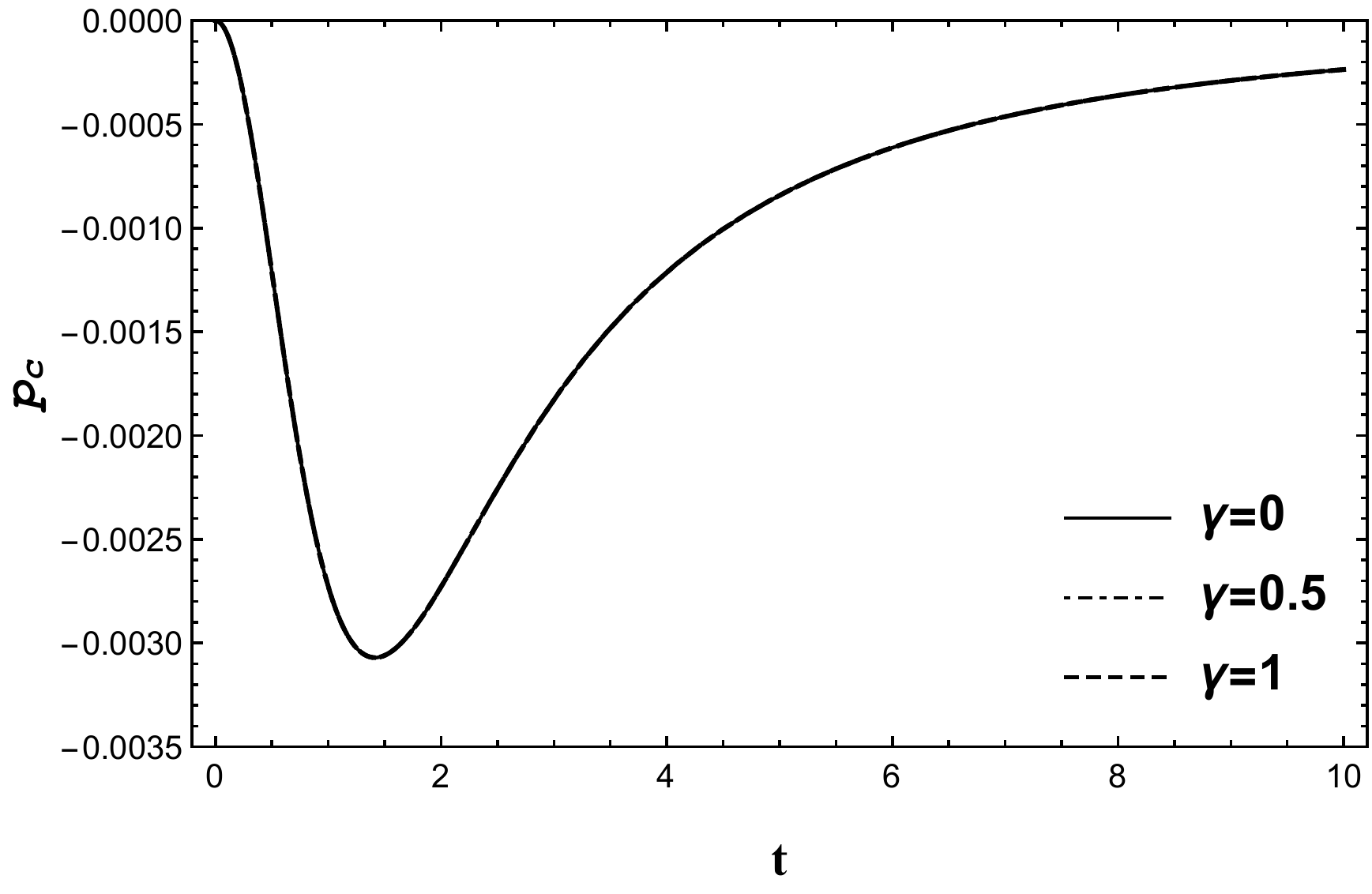} \\ 
\mbox (a) & \mbox (b) & \mbox (c)%
\end{array}%
$%
\end{center}
\caption{ \textit{The plots of }$\protect\rho $\textit{, }$p_{m}$\textit{\
and }$p_{c}$\textit{\ Vs. time }$t$\textit{\ for the Model-III with }$%
\mathit{A=}\frac{1}{3}$\textit{, }$\mathit{B=1}$\textit{, }$\protect\lambda %
=1$\textit{, }$\protect\eta =0.25$\textit{\ and }$\protect\alpha =2$ \& $%
\protect\beta =\frac{1}{2}$\textit{\ for the particular values of }$\protect%
\gamma =0$, $0.5$, $1$\textit{.}}
\end{figure}

Fig. 3a illustrate that the energy density $\rho $ initiates with a finite
value, increases promptly for a short period of time, attains its maximum
then goes down and gradually decreases with slow rate and remains finite
forever. In Fig. 3b, we see the matter pressure $p_{m}$ begins with a finite
negative value, increases in a small interval of time but remains negative
then drops again and start decreasing and eventually remains negative
forever. Fig. 3c depicts that initially there is no matter creation, as time
evolves particles get created and pressure $p_{c}$ starts decreasing
promptly in a short span of time, then it raised up, expands as time derive,
remains negative then tends to zero as $t\rightarrow \infty $. Here in this
model, we observe that the universe is accelerated expanding since $p^{c} $
is negative. The rate of particle production are initially negligible, and
again increases to attain high rate of particle production, and then
gradually decreases at late time.

\section{Statefinder diagnostics}

\noindent \qquad \qquad In order to describe various dark energy models a
method have been developed in \cite{Sahni1} known as statefinder diagnostic.
The statefinder diagnostic pair $\{r,s\}$ is a geometrical parameter that
probe the expansion dynamics of the Universe through the higher derivatives
of the scale factor $a$ \textit{i.e.}, Hubble parameter $H$ and the
deceleration parameter $q$, and is defined by%
\begin{equation}
r=\frac{\dddot{a}}{aH^{3}}\text{, \ \ }s=\frac{r-1}{3(q-\frac{1}{2})},
\label{24}
\end{equation}%
where $q\neq \frac{1}{2}$. For different dark energy models, the
trajectories in the $s-r$ plane show diverge behavior. For example, in a
spatially flat FRW Universe, the $\Lambda CDM$ model is identified as a
fixed point $\{0,1\}$ in the $\{s,r\}$ diagram while the standard $SCDM$
model corresponds to a fixed point $\{1,1\}$ in the $\{s,r\}$ diagram. This
analysis can successfully differentiate quintessence, Chaplygin gas,
braneworld dark energy models and some other interacting dark energy models.
For some particular model, the position of the fixed point $\{s,r\}$ can be
calculated and located in the diagram. The statefinder diagnostics for
different dark energy models have been discussed in references \cite{sami1,
sami2, sami3,}.

Here, we have obtained three different dark energy models that shows quite
different evolution. We apply the statefinder diagnostic technique to
calculate the diverging or converging behavior of our obtained dark energy
models with respect to the $SCDM$ or $\Lambda CDM$ model. The $r,s$
parameters for our models-I are obtained as 
\begin{equation}
r=1-\frac{3}{\beta }+\frac{2}{\beta ^{2}},\text{ }s=\frac{2}{3\beta }.
\label{25}
\end{equation}

Similarly, for model-II and model-III, the $\{r,s\}$ pair are obtained as%
\begin{eqnarray}
r &=&1+\left( -\frac{3\alpha }{\beta }+\frac{2\alpha ^{2}}{\beta ^{2}}%
\right) +\left( -\frac{6}{\beta }+\frac{6\alpha }{\beta ^{2}}\right) t+\frac{%
6}{\beta ^{2}}t^{2},  \notag  \label{26} \\
s &=&\frac{2\alpha -\beta }{4\beta }+\frac{1}{\beta }t+\frac{4\alpha
^{2}-9\beta ^{2}}{12\beta (2\alpha -3\beta +4t)},
\end{eqnarray}%
and 
\begin{eqnarray}
r &=&\frac{(3\alpha +(\beta -1)t^{2})(\beta -2)}{t^{2}\beta ^{2}},  \notag
\label{27} \\
s &=&\frac{-6\alpha (\beta -2)+(6\beta -4)t^{2}}{6\alpha \beta +3\beta
(3\beta -2)t^{2}},
\end{eqnarray}%
respectively. We can see, in the power law model (model-I) $q$, $r$, $s$ are
constants and has only one model parameter $\beta $. For different values of 
$\beta $, we have different expansion factors and can be analyzed in the
following Table 1.

{\scriptsize 
\begin{table}[tbh]
\begin{center}
{\scriptsize \textbf{Table 1.}\vskip0.1in 
\begin{tabular}{|c|c|c|c|c|c|c}
\cline{1-6}
$\beta $ & $a(t)$ & $H(t)$ & $q$ & $r$ & $s$ &  \\ \cline{1-6}
$2$ & $ct^{2}$ & $\frac{2}{t}$ & $-0.5$ & $0$ & $0.33$ &  \\ \cline{1-6}
$\frac{3}{2}$ & $ct^{\frac{3}{2}}$ & $\frac{3}{2t}$ & $-0.33$ & $-0.11$ & $%
0.44$ &  \\ \cline{1-6}
$\frac{2}{3}$ & $ct^{\frac{2}{3}}$ & $\frac{2}{3t}$ & $0.5$ & $1$ & $1$ & $%
SCDM $ \\ \cline{1-6}
$\infty $ & $ct$ & $0$ & $-1$ & $1$ & $0$ & $\Lambda CDM $ \\ \cline{1-6}
\end{tabular}
\label{table:nonlin}  }
\end{center}
\end{table}
} In the model-II and the model-III, the parameters $q$, $r$, $s$ are time
varying and contains two model parameters $\alpha $ and $\beta $. To have a
better understanding of our model behavior, we plot the trajectories of the
models in $s-r$ and $q-r$ planes.

\begin{figure}[htbp]
\begin{center}
$%
\begin{array}{c@{\hspace{0.1 in}}cc}
\includegraphics[width=2.5 in]{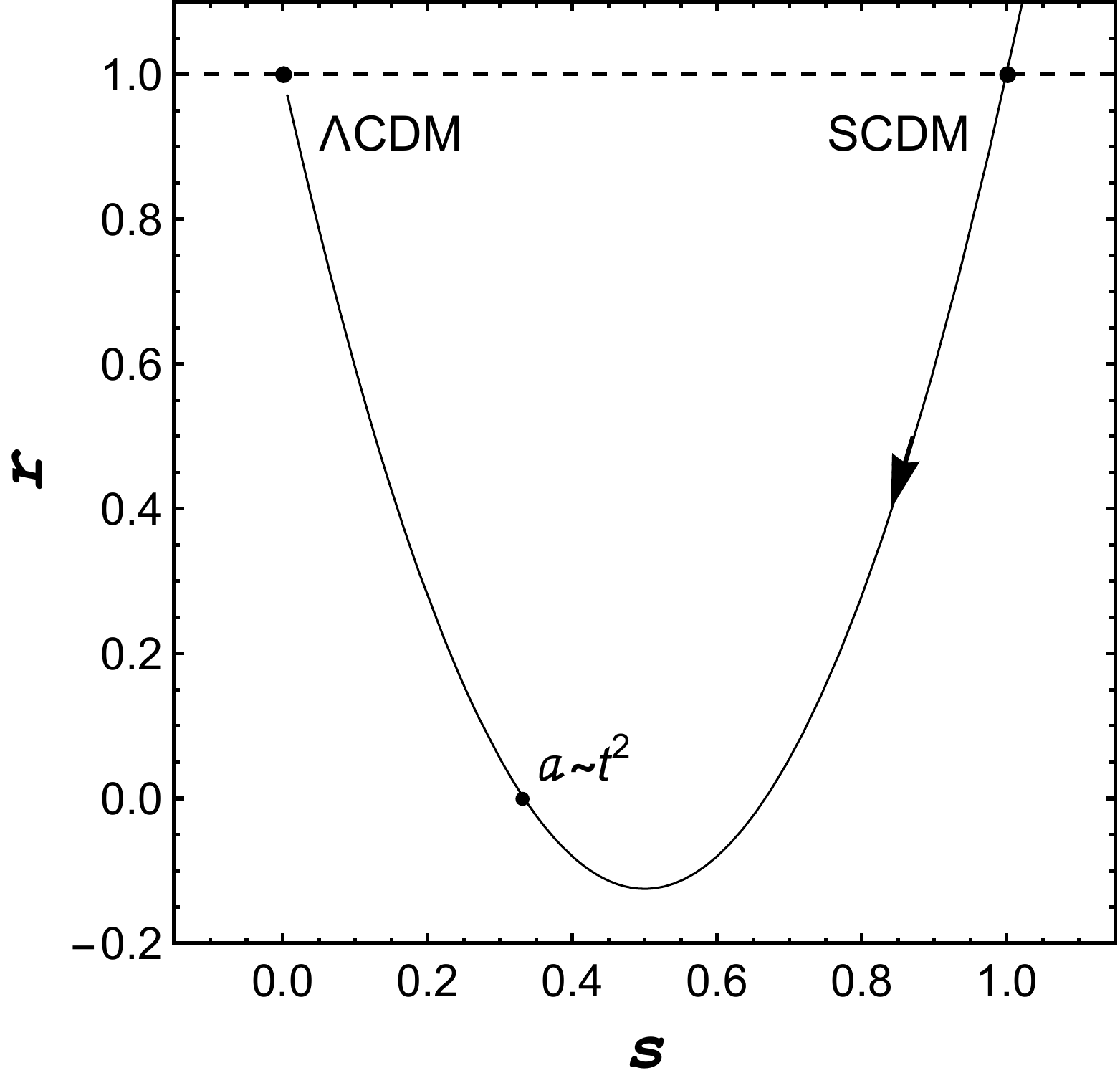} & 
\includegraphics[width=2.5 in,
height=2.4 in]{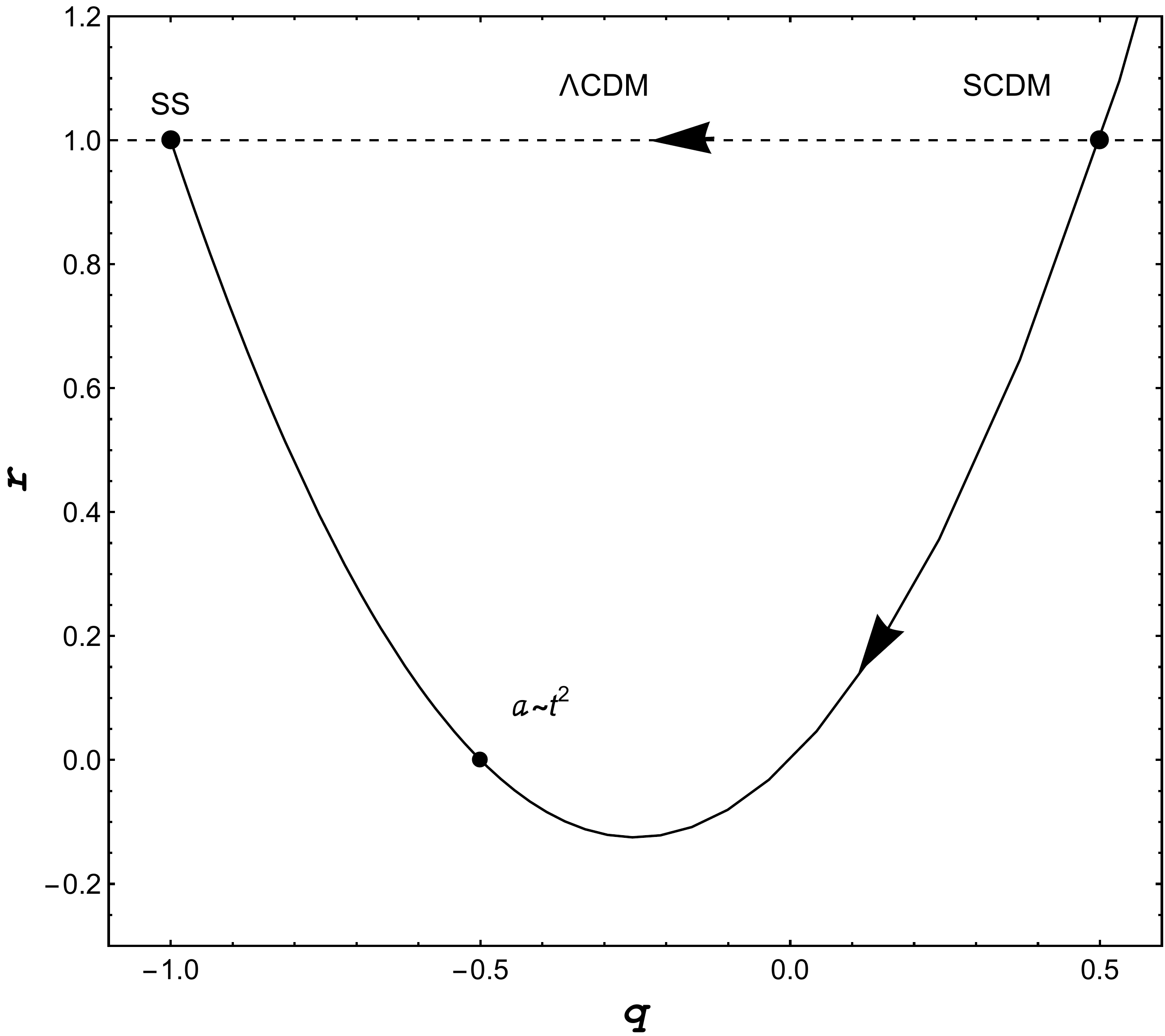} &  \\ 
\mbox (a) & \mbox (b) & 
\end{array}%
$%
\end{center}
\caption{\textit{\ The }$\mathit{s-r}$\textit{\ and }$\mathit{q-r}$\textit{\
diagrams in Model-I.}}
\end{figure}

The left panel in Fig. 4 depicts the evolution of trajectory for different $%
\beta $ in $s-r$ plane. Initially the trajectory evolves and converges to
the fixed point $SCDM $ ($s=1$, $r=1$). The value of both `$r$' and `$s$'
start decline and attains their minimum value, after that both `$r$' and `$s$%
' increases towards the fixed point $\Lambda CDM $ ($s=0$, $r=1$). The
horizontal line in the above diagram shows the progression of model from the 
$SCDM $ to $\Lambda CDM $ for increasing values of $\beta $. The right panel
in Fig. 4 shows the evolution of trajectory in the $q-r$ plane. The
trajectory in the $q-r$ plane behaves alike as in the $s-r$ plane but here
in the $q-r$ plane, we can observe that the model converges to the steady
state model $(SS) $ ($q=-1$, $r=1$) and transform from the fixed point $SCDM 
$ ($q=\frac{1}{2}$, $r=1$) to $\Lambda CDM $ and end up with the fixed point 
$SS $ with increasing values of $\beta $. A particular model for $\beta =2$ 
\textit{i.e.}, $a\sim t^{2}$ is shown in the figures.

\begin{figure}[tbph]
\begin{center}
$%
\begin{array}{c@{\hspace{.1in}}cc}
\includegraphics[width=2.5 in]{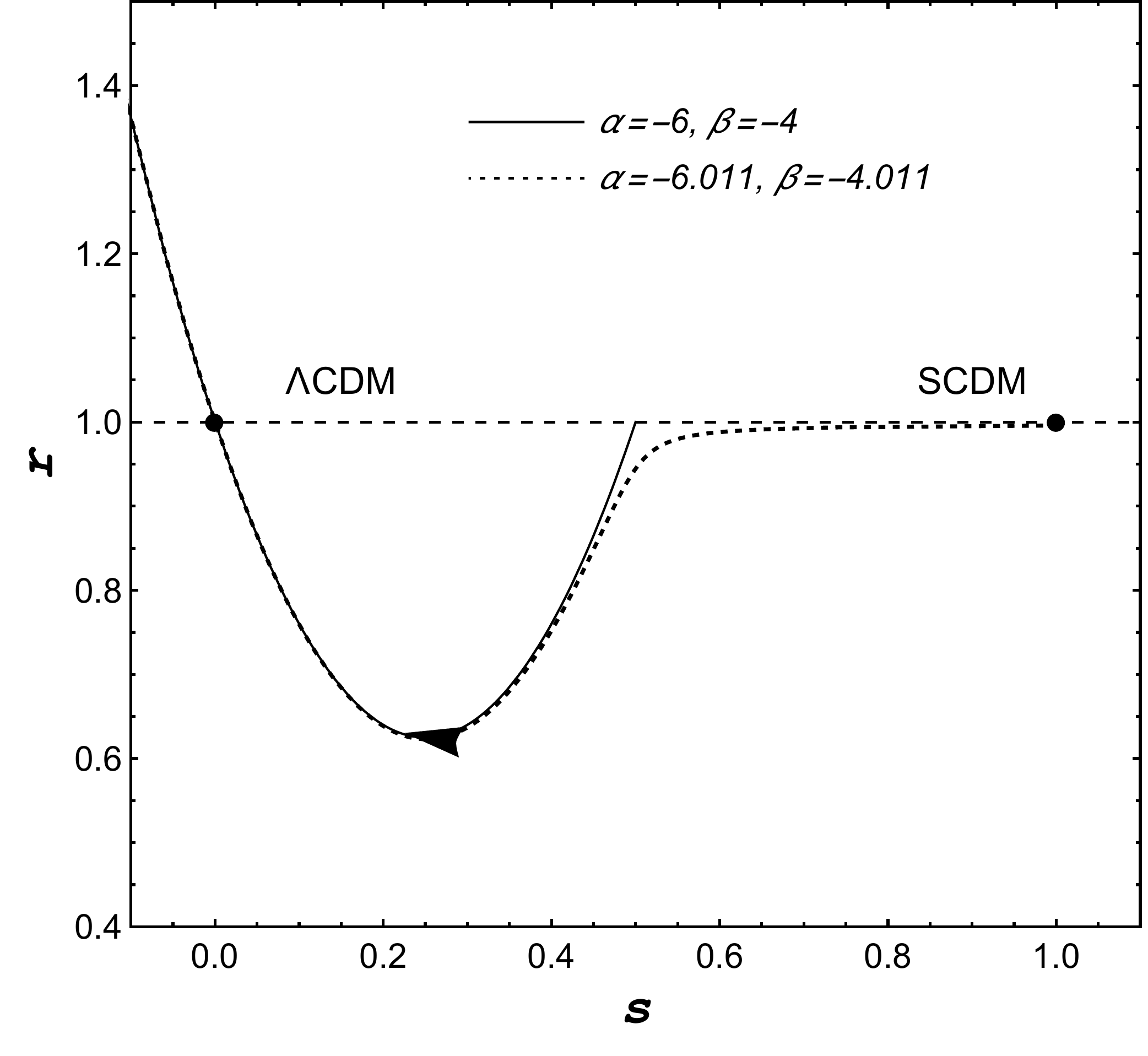} & 
\includegraphics[width=2.5 in,
height=2.4 in]{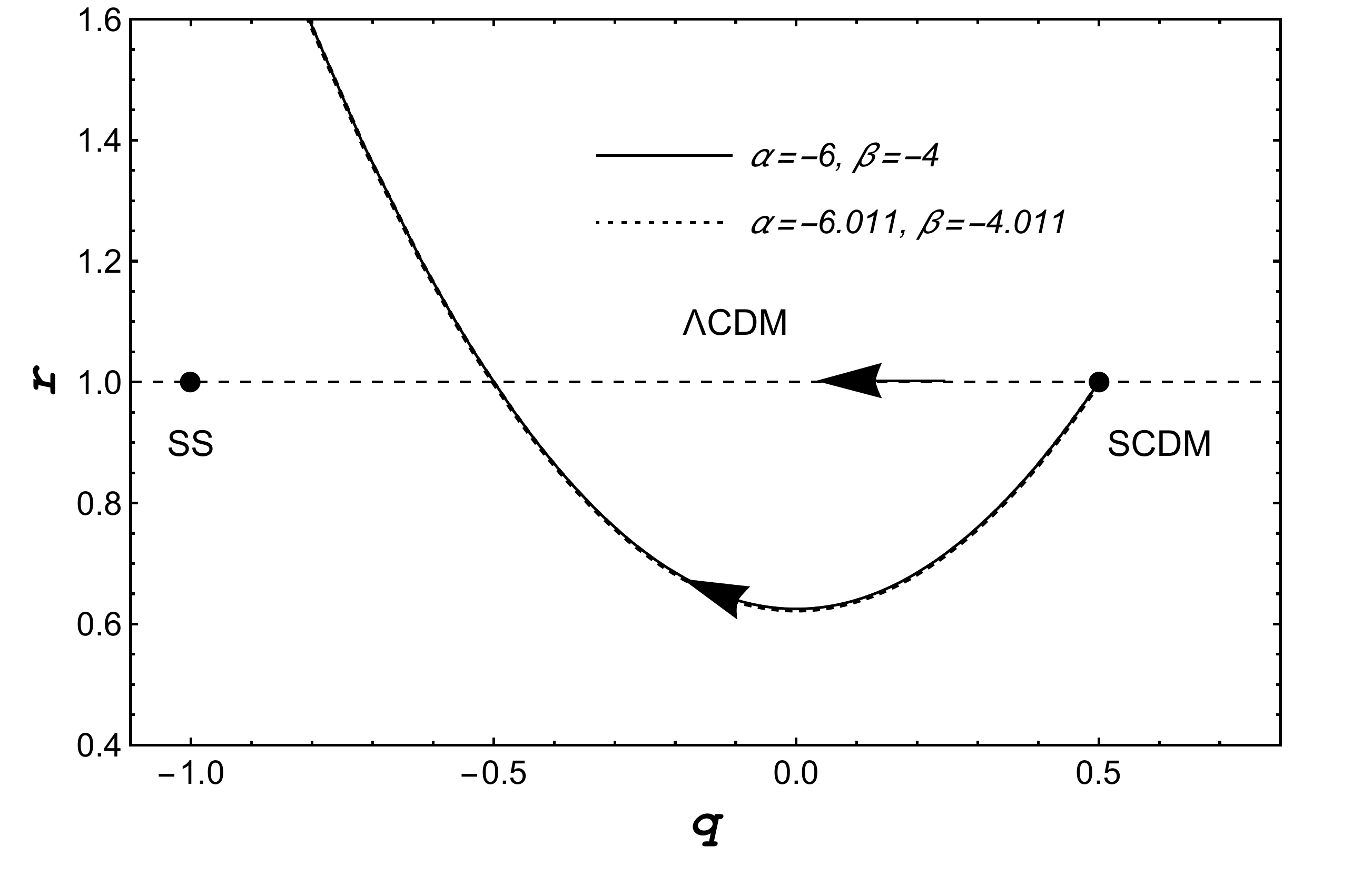} &  \\ 
\mbox (a) & \mbox (b) & 
\end{array}%
$%
\end{center}
\caption{\textit{\ The }$\mathit{s-r}$\textit{\ and }$\mathit{q-r}$\textit{\
diagrams in Model-II.}}
\end{figure}

For model-II, the trajectories in the $s-r$ and $q-r$ planes are shown in
Fig. 5. The left panel in Fig. 5 displays the evolution of trajectory with
time in the $s-r$ diagram. The statefinder ($s-r$) is shown for two
different values of model parameters $\alpha $ and $\beta $. The solid black
trajectory does not begins with $SCDM$, evolves with time and touches the $%
\Lambda CDM$. A slightly change in the value of model parameters gives rise
to a new trajectory (dashed line) which starts with $SCDM$, advanced with
time and eventually coincident with the other curve and ultimately converges
to $\Lambda CDM$. The right panel in Fig. 5 depicts the trajectories in $q-r$
plane for two different values of model parameters $\alpha $, $\beta $ which
evolves with time. Both the trajectory initially start with the $SCDM$ but
never touches to the $SS$ model.

\begin{figure}[tbph]
\begin{center}
$%
\begin{array}{c@{\hspace{.1in}}cc}
\includegraphics[width=2.5 in]{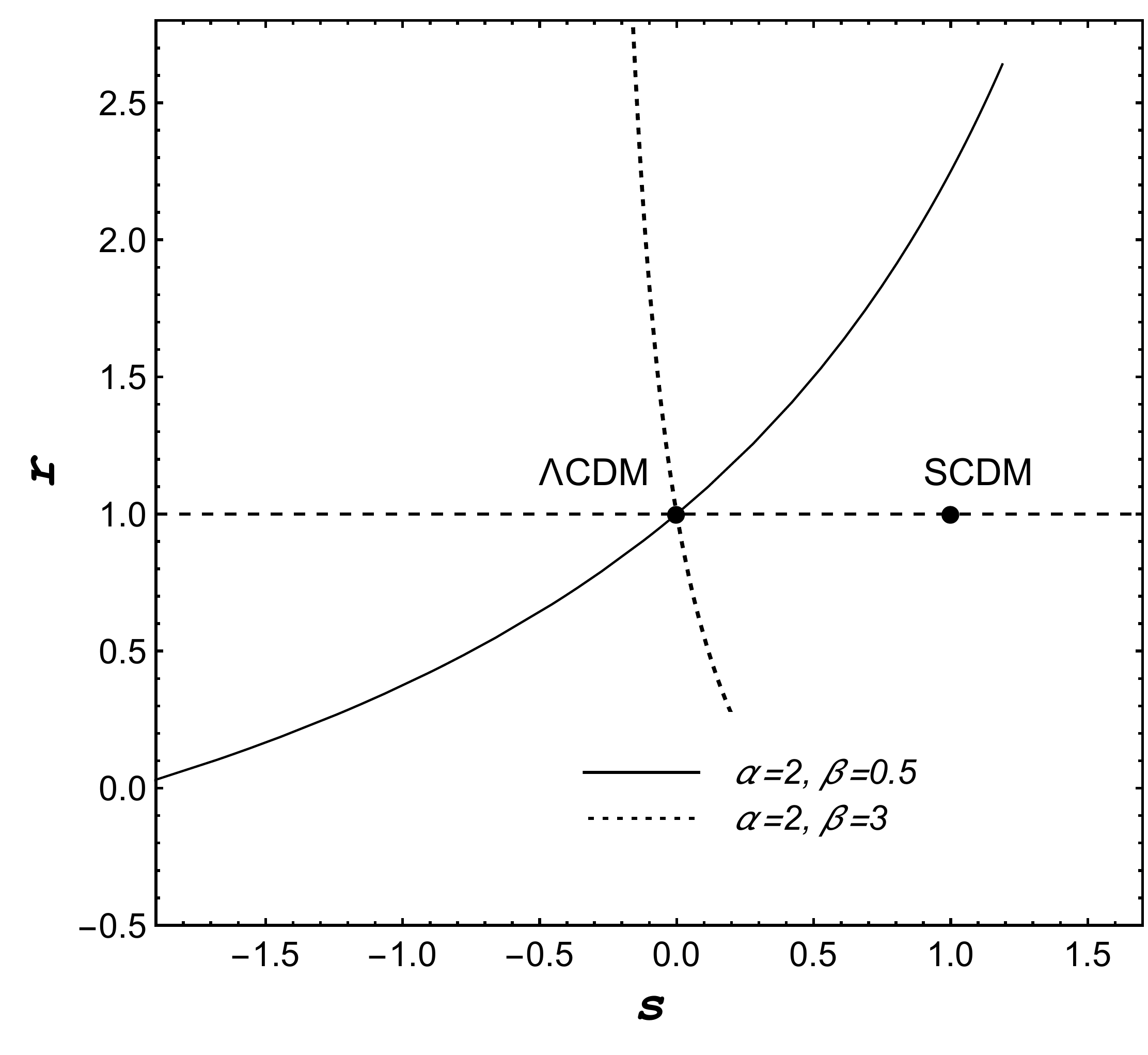} & 
\includegraphics[width=2.5 in,
height=2.3 in]{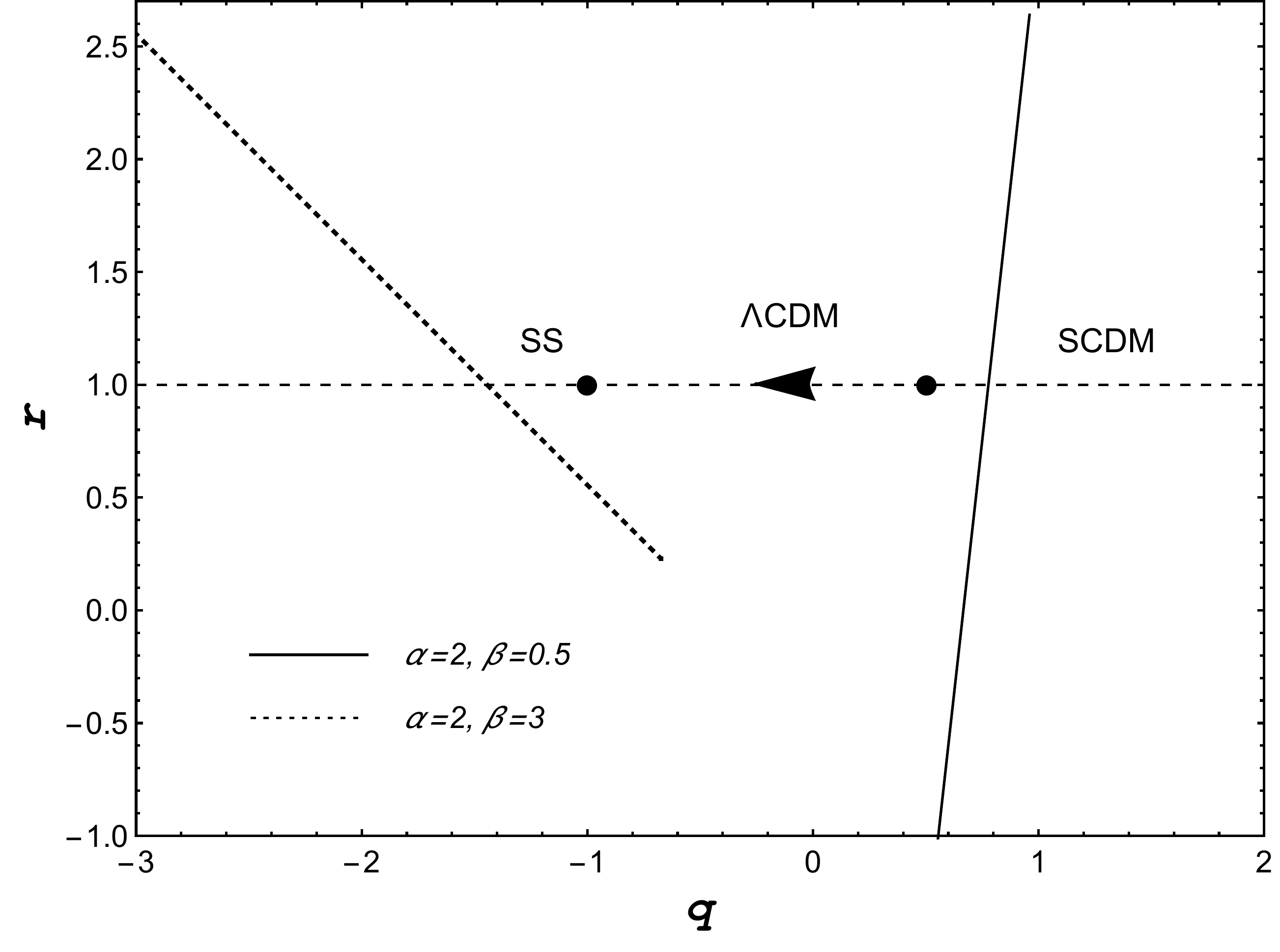} &  \\ 
\mbox (a) & \mbox (b) & 
\end{array}%
$%
\end{center}
\caption{\textit{\ The }$\mathit{s-r}$\textit{\ and }$\mathit{q-r}$\textit{\
diagrams in Model-III.}}
\end{figure}
\noindent \qquad

Finally, for model-III, the trajectories in the $s-r$ and $q-r$ planes are
shown in Fig. 6. The left panel in Fig. 6 in the above figure displays the
evolution of trajectory with time in the $s-r$ diagram. The statefinder ($%
s-r $) is shown for two different values of model parameters $\alpha $ and $%
\beta $. The solid black trajectory does not begins with $SCDM$, evolves
with time and touches the $\Lambda CDM$. A change in the value of model
parameters $\beta $ from $0.5$ to $3$ gives rise to a new trajectory (black
dashed) which also deviates from $SCDM$ and intersect the point $\Lambda CDM$%
, plane for two different values of model parameters $\alpha $, $\beta $
which evolves with time. Both the trajectory deviates from $SCDM$ and $SS$
model.

\section{Velocity of Sound and stability of model}

\noindent \qquad \qquad As we know, the stability of linear perturbations is
a critical test for the viability of any cosmological model which is beyond
our scope here. However, a stringent constraint comes from imposing the
Velocity of sound ($C_{s}^{2}$) to be sufficiently smaller than $1$ to avoid
unwanted oscillations in the matter power spectrum. We plot $C_{s}^{2}$ for
our obtained models with suitable choice of the parameters involved and are
shown in the following figure: 
\begin{figure}[tbph]
\begin{center}
$%
\begin{array}{c@{\hspace{.1in}}cc}
\includegraphics[width=2.2 in]{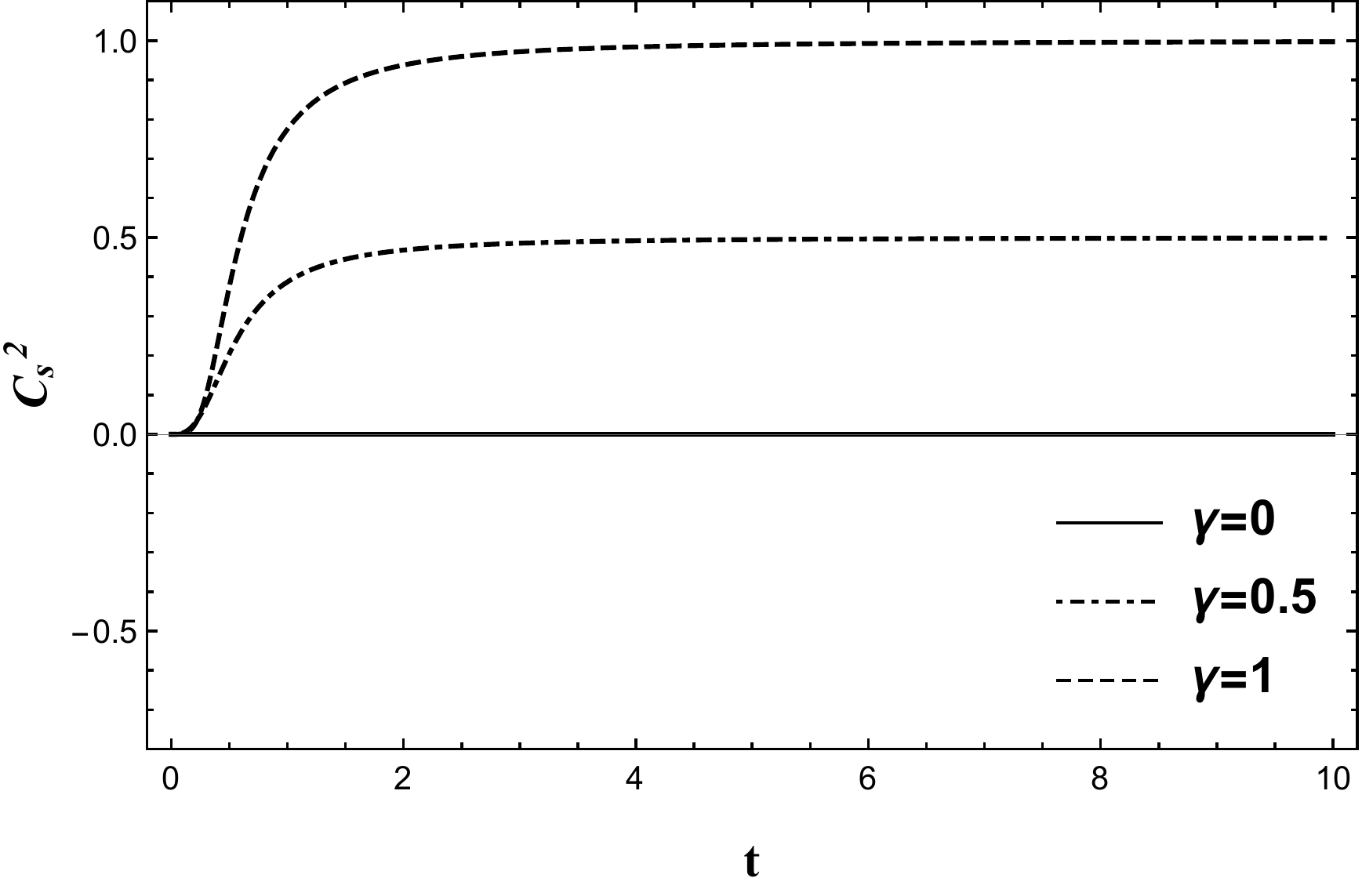} & 
\includegraphics[width=2.2
in]{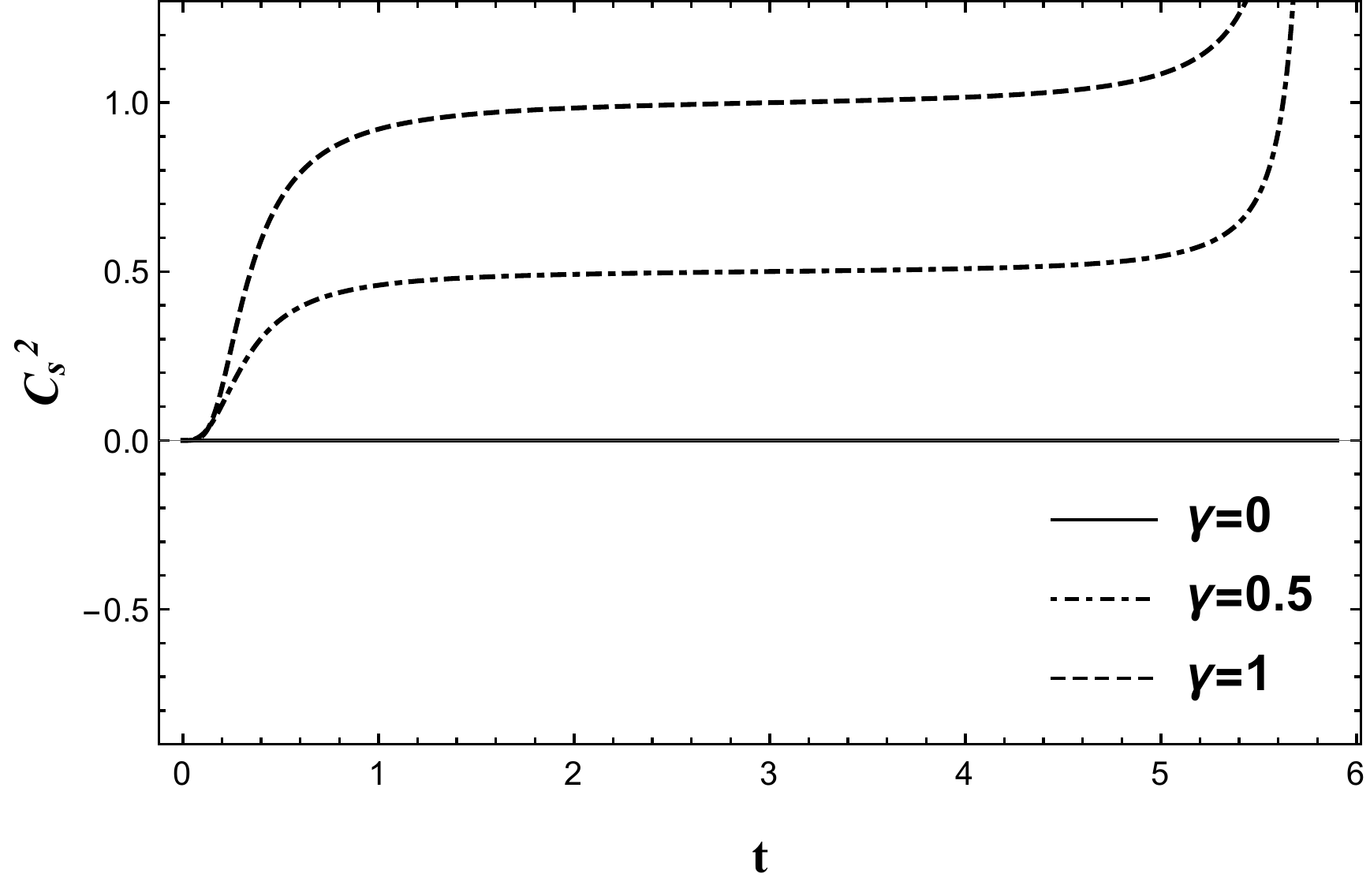} & \includegraphics[width=2.2 in]{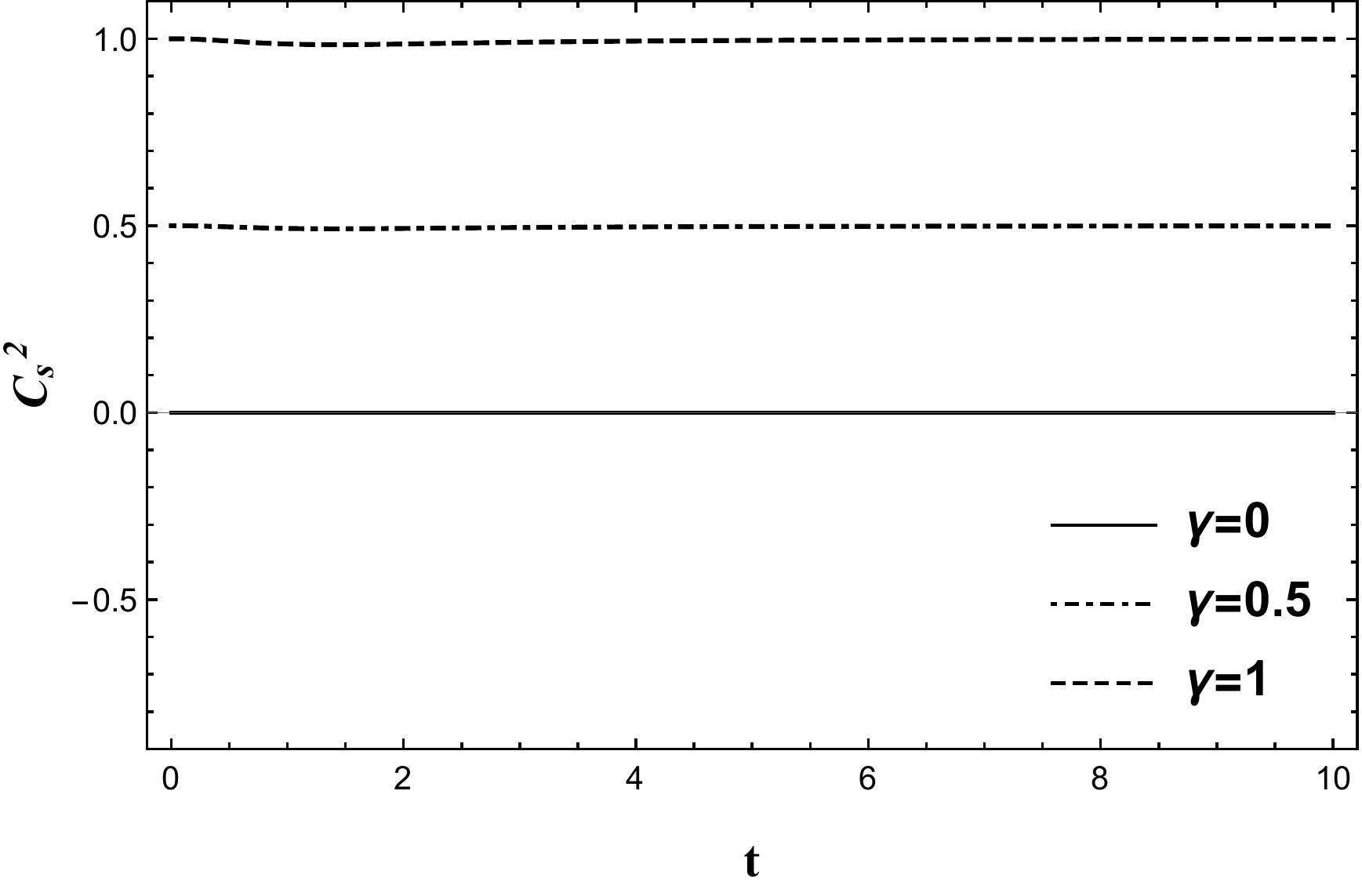} \\ 
\mbox (a) & \mbox (b) & \mbox (c)%
\end{array}%
$%
\end{center}
\caption{\textit{\ The plots of velocity of sound }$C_{s}^{2}$\textit{\ Vs.
time }$t$\textit{\ in Model-I, II and III respectively for }$\mathit{A=}$%
\textit{\ }$\frac{1}{3}$\textit{, }$\mathit{B=1}$\textit{\ and }$\protect%
\eta $\textit{\ }$\mathit{=0.25}$\textit{\ for some selected value of }$%
\protect\gamma $\textit{.}}
\end{figure}

Fig. 7 depicts the stability of our obtained models for the restricted
values of the model parameters $\alpha $ and $\beta $ and the chosen
particular values of other constants.

\section{Some kinematic behavior}

\subsection{Lookback time}

\qquad The lookback time $t_{L}$ to an object is the time elapsed between
the detection of light today $(z=0)$ and at the time of emission of photons
at a particular redshift $z$.\newline
\begin{equation}
t_{L}=t_{0}-t(z)=\int_{a}^{a_{0}}\frac{dt}{\dot{a}},  \label{28}
\end{equation}%
where $a_{0}$ indicates the value of scale factor $a(t)$ at present time $%
t_{0}$, and by the relationship between the scale factor and redshift $z$ 
\begin{equation}
a_{0}=a(t)(1+z),  \label{29} \\
\end{equation}

Here, for Model-I, II and III, the $t-z$ relation takes the form 
\begin{equation}
t(z)=\beta H_{0}^{-1}(1+z)^{-\frac{1}{\beta }},  \label{30}
\end{equation}%
\begin{equation}
t(z)=\frac{\alpha }{(1+z)^{\frac{\alpha }{\beta }}\big(1+\frac{\alpha
(\alpha +1)H_{0}}{\beta }\big)-1},  \label{31}
\end{equation}%
\begin{equation}
t(z)=\frac{\beta H_{0}^{-1}}{1+\alpha }\left[ (1+z)^{\frac{-2}{\beta }%
}(1+\alpha )-\alpha \right] ^{\frac{1}{2}},  \label{32}
\end{equation}%
\newline
where $H_{0}$ is the present day Hubble parameter of the Universe, $\alpha $
and $\beta $ are the model parameters.

\subsection{Proper distance}

The proper distance between two events is the distance between them in the
frame of reference in which they occur at exactly same time and measured by
a ruler at the time of observation. Proper distance is defined as $%
d(z)=a_{0}r$, where $r=r(z)$ is the radial distance of the object, which is
given by 
\begin{equation}
r(z)=\int_{t}^{t_{0}}\frac{dt}{a(t)},  \label{33}
\end{equation}

For the above discussed Model I, II, III, the proper distance d(z) are given
respectively: 
\begin{equation}
d(z)=\frac{H_{0}^{-1}}{(\frac{1}{\beta }-1)}[1-(1+z)^{-(\frac{1}{\beta }%
-1)}],  \label{34}
\end{equation}%
\newline
\begin{equation}
d(z)=a_{0}\Big[\frac{\alpha t}{c(\alpha -\beta )}\big(\frac{t}{\alpha }\big)%
^{-\frac{\beta }{\alpha }}\times Hypergeometric\,2F1[-\frac{\beta }{\alpha }%
,1-\frac{\beta }{\alpha },2-\frac{\beta }{\alpha },-\frac{t}{\alpha }]\Big]%
_{t}^{t_{0}},  \label{35}
\end{equation}%
\newline
\begin{equation}
d(z)=a_{0}\Big[\Big(\frac{t(t^{2}+\alpha )^{1-\frac{\beta }{2}}}{c\alpha }%
\Big)\times Hypergeometric\,2F1[1,\frac{3-\beta }{2},\frac{3}{2},-\frac{t^{2}%
}{\alpha }]\Big]_{t}^{t_{0}}.  \label{36}
\end{equation}%
\newline

Here, in order to make the plots, we have considered the series of the above
mentioned Hypergeometric functions upto third order term.\newline
\begin{figure}[tbph]
\begin{center}
$%
\begin{array}{c@{\hspace{.1in}}cc}
\includegraphics[width=2.5 in, height=2.0 in]{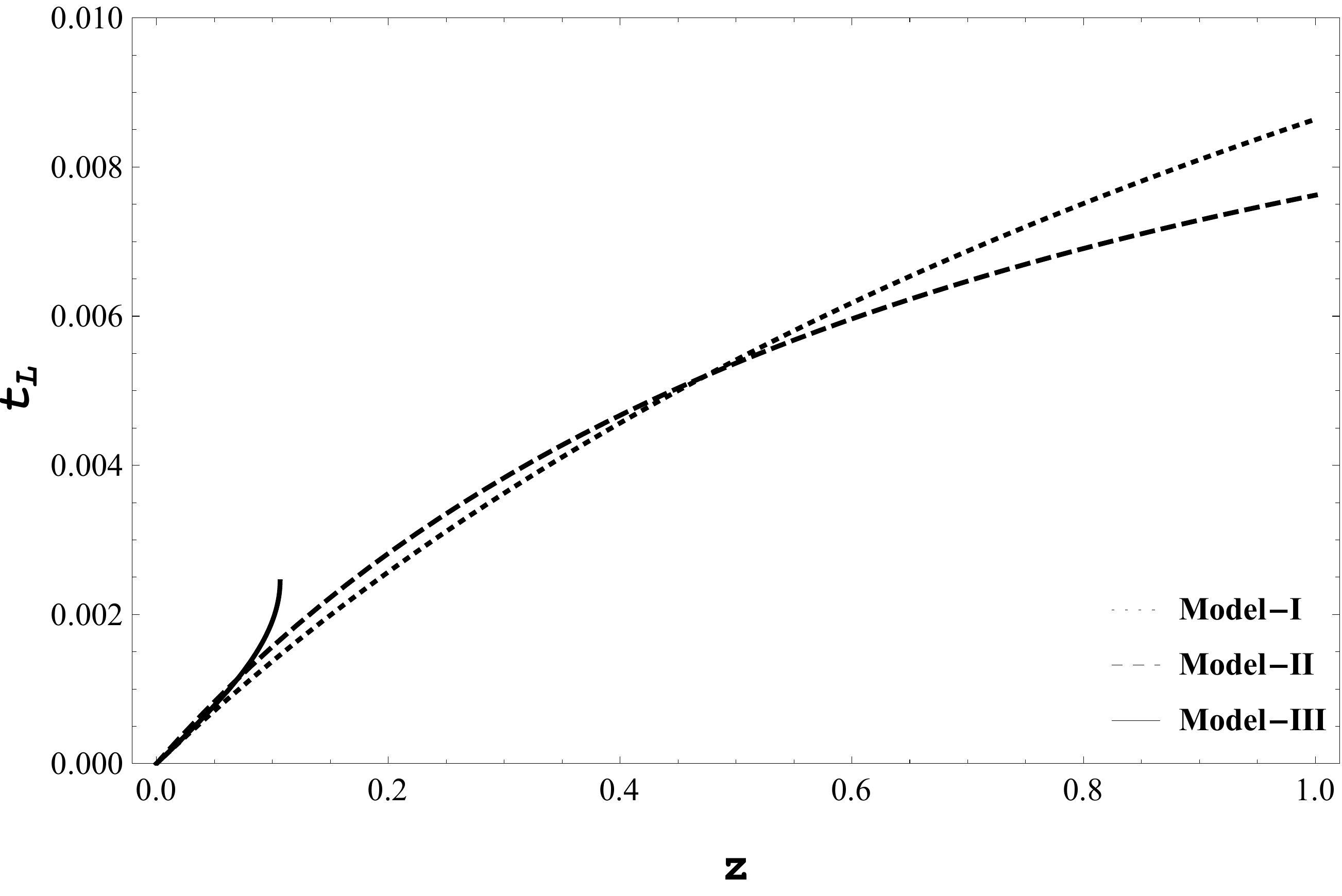} & %
\includegraphics[width=2.5 in, height=2.0 in]{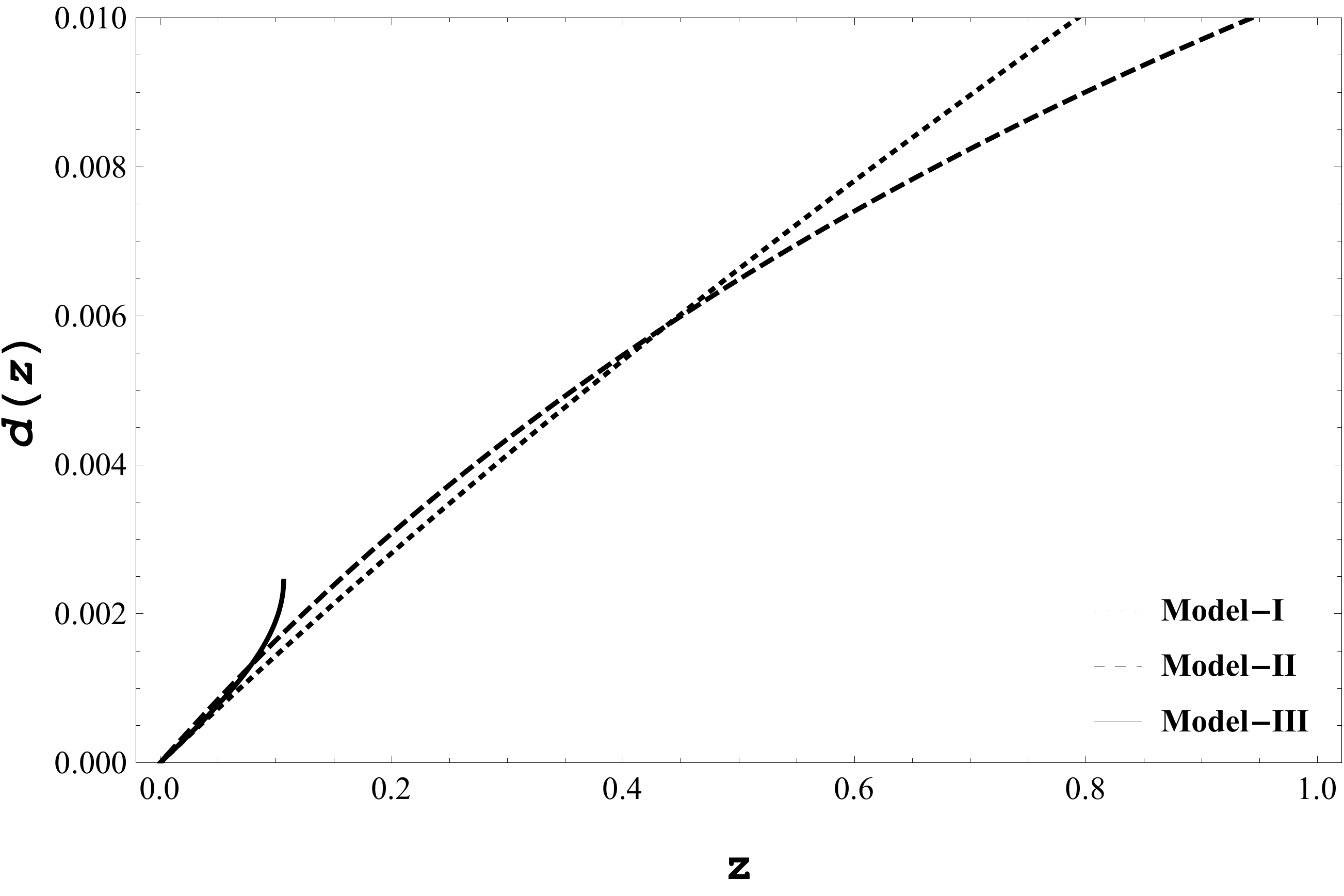} &  \\ 
\mbox (a) & \mbox (b) & 
\end{array}%
$%
\end{center}
\caption{\textit{\ The plots of look back time }$t_{L}$ \textit{and proper
distance} $d(z)${\textit{\ Vs. redshift} }$z$\textit{\ for the Model-I, II,
III.}}
\end{figure}

\subsection{Luminosity distance}

Luminosity distance $d_{l}$ of a source with redshift $z$ is defined by the
relation 
\begin{equation}  \label{37a}
d_{l}^{2}=\frac{l}{4\pi L},
\end{equation}%
\newline
where $L$ is the flux measured and $l$ is the luminosity of the object. From
equation (\ref{33}) the luminosity distance is given by 
\begin{equation}  \label{37}
d_{l}=(1+z)d(z).
\end{equation}

\subsection{Angular diameter distance}

The angular diameter distance is defined by 
\begin{equation}  \label{38}
\displaystyle d_{A}={\frac{l_{1}}{\theta }} ,
\end{equation}
where {$\displaystyle l_{1}$} is a physical size and {$\displaystyle\theta $ 
} is the angular size of an object, and the angular diameter distance $d_{A}$
of an object in terms of redshift $z$ is 
\begin{equation}  \label{39}
d_{A}=\frac{d(z)}{1+z}=\frac{d_{l}}{(1+z)^{2}}.
\end{equation}

\begin{figure}[tbph]
\begin{center}
$%
\begin{array}{c@{\hspace{.1in}}cc}
\includegraphics[width=2.5 in, height=2.0 in]{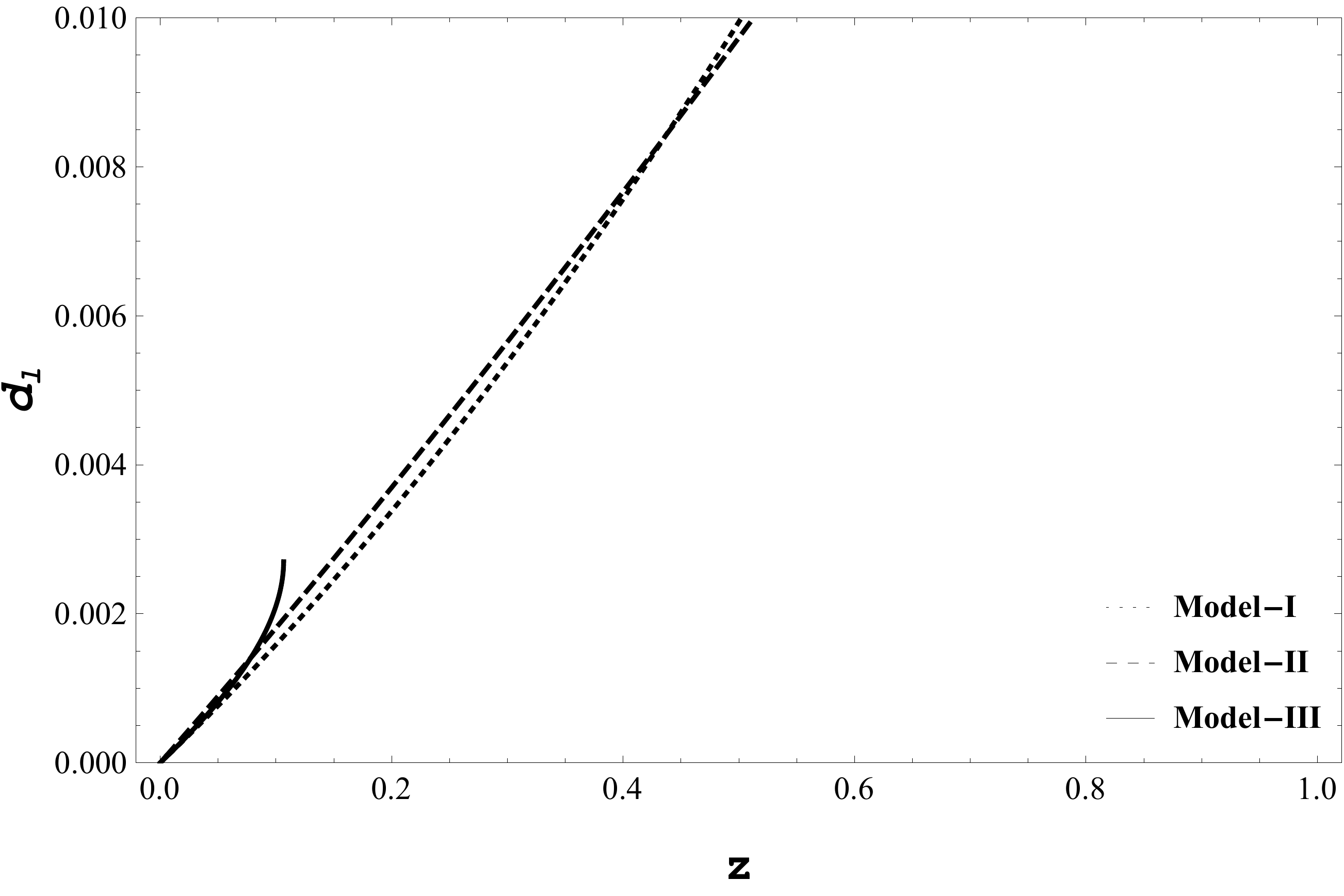} & %
\includegraphics[width=2.5 in, height=2.0 in]{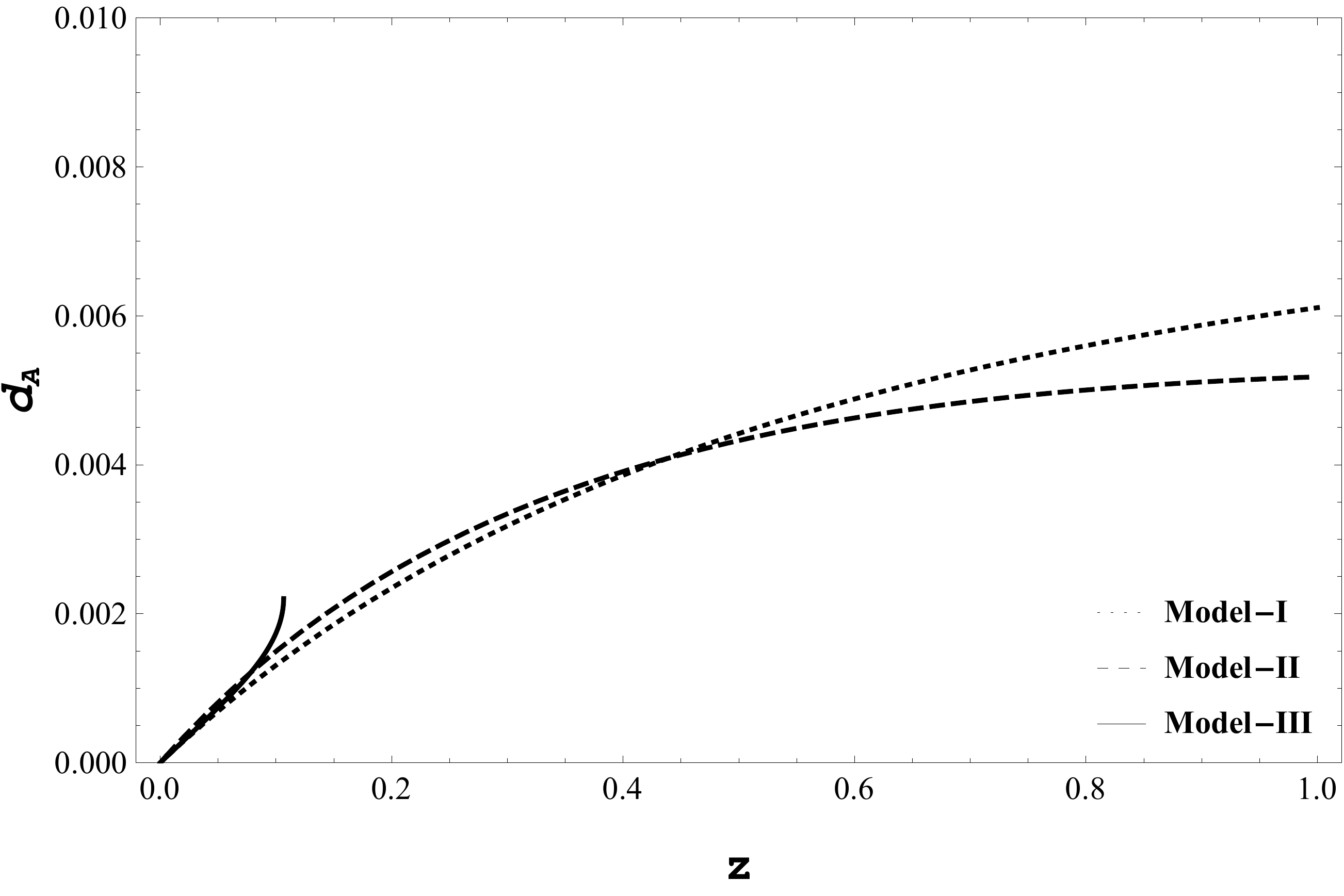} &  \\ 
\mbox (a) & \mbox (b) & 
\end{array}%
$%
\end{center}
\caption{\textit{\ The plots of luminosity distance }$d_{l}$ \textit{and
angular diameter distance} $d_{A}${\textit{\ Vs. redshift} }$z$\textit{\ for
the Model-I, II, III.}}
\end{figure}

\subsection{Deceleration parameter and phase transition}

\noindent \qquad \qquad The deceleration parameter for model-I is constant
throughout the evolution. For model-I, $q<0$ for $\beta >1$ and $q>0$ for $%
\beta <1$. Deceleration parameter for model-II and model-III is time varying
and can be rewritten in terms of redshift $z$ using $t-z$ relationship (\ref%
{31}) and (\ref{32}) as $q(z)=-1+\frac{\alpha }{\beta }+\frac{2}{\beta }%
\frac{\alpha }{-1+(1+\alpha )(1+z)^{\frac{\alpha }{\beta }}}$ for model-II
and $q(z)=-1+\frac{1}{\beta }-\frac{\alpha }{\beta }\frac{1}{-\alpha
+(1+\alpha )(1+z)^{-\frac{2}{\beta }}}$ for model-III and can be analyzed in
the following Table 2.\newline

\begin{center}
{\scriptsize \textbf{Table 2.}\vskip0.1in 
\begin{tabular}{|c|c|c|c|}
\hline
redshift & DP & Value of DP, model-II & Value of DP, model-III \\ \hline
$z=\infty $ & $q_{i}$ & $-1+\frac{\alpha }{\beta }$ & $-1+\frac{2}{\beta }$
\\ \hline
$z=0$ & $q_{0}$ & $-1+\frac{2+\alpha }{\beta }$ & $-1+\frac{1-\alpha }{\beta}
$ \\ \hline
$z=-1$ & $q_{f}$ & $-1-\frac{\alpha }{\beta }$ & $-1+\frac{1}{\beta }$ \\ 
\hline
\end{tabular}%
}
\end{center}

Here $q_{i}$ is the initial value of DP at the time of big bang, $q_{0}$
being the present value of the DP and $q_{f}$ is the value of DP in the
infinite future. The model parameters $\alpha $ and $\beta $ are to be
chosen carefully so that we can have $q_{0}<0$ explaining the observation
along with the positivity condition of the energy density $\rho $. For the
appropriate values of $\alpha $ and $\beta $, model-II can exhibit a phase
transition from deceleration to acceleration while model-III will show
acceleration to deceleration phase transition or eternal acceleration. One
can also constrain the values of $\alpha $ and $\beta $ through any
observational data which will be defer to our future investigation. As the
present observation strongly reveals a phase transition from deceleration to
acceleration in the near past and our obtained model-II fits well in this
context, we can plot a graph (see Fig. 10) showing the phase transition
redshift ($z_{tr}$) and the present value of deceleration parameter ($q_{0}$%
).

\begin{figure}[tbph]
\begin{center}
\includegraphics[width=3.5 in, height=2.5 in]{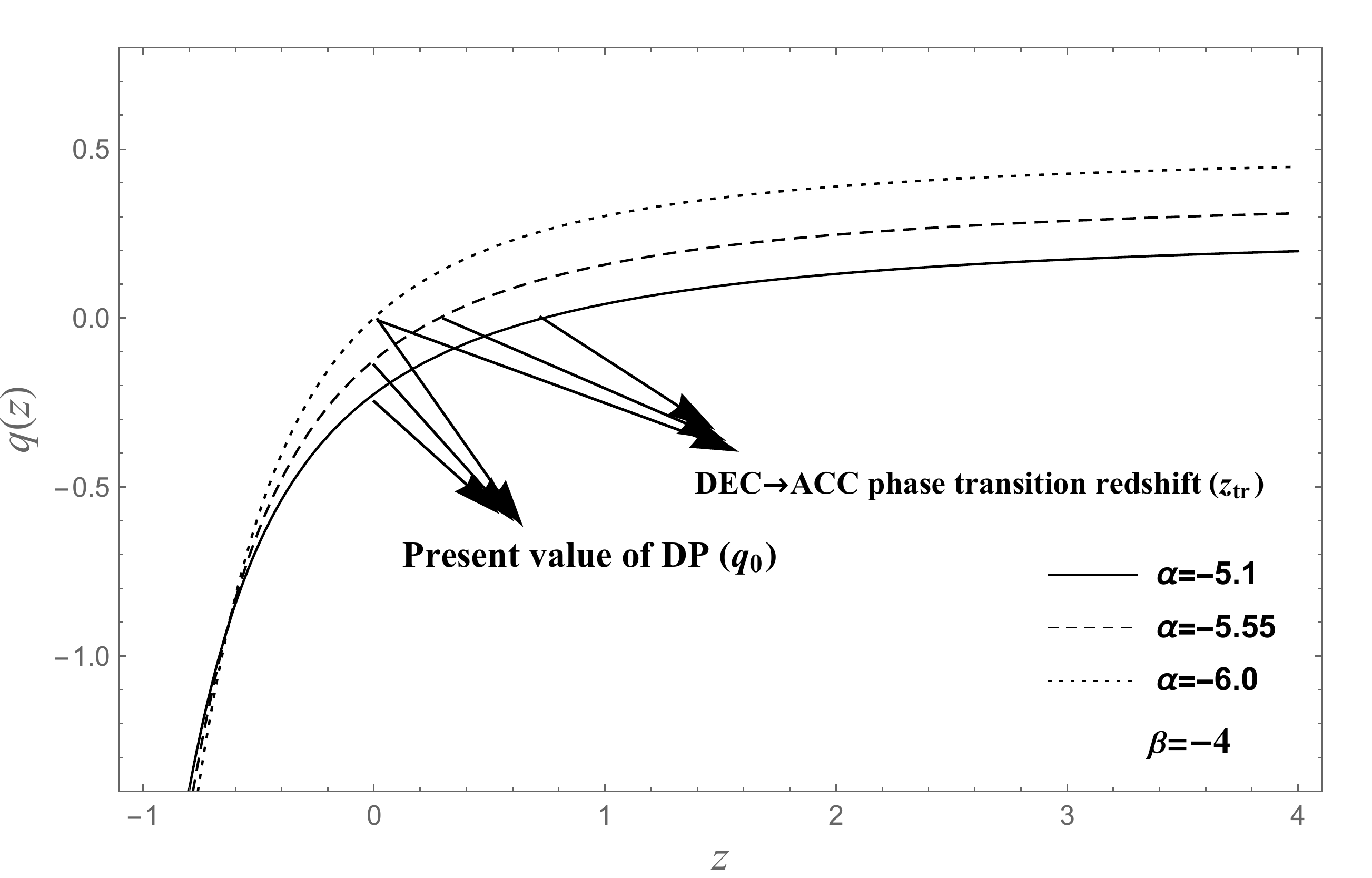}
\end{center}
\caption{\textit{\ The plot shows the deceleration to acceleration phase
transition redshift ($z_{tr}$) and the present value of deceleration
parameter ($q_0$) for model-II for fixed $\protect\beta=-4.0$ and different $%
\protect\alpha=-5.1, -5.55, -6.0$.}}
\end{figure}
\newpage

\section{Discussion and Conclusion}

\noindent \qquad \qquad In this paper, we explored the FLRW model with
modified Chaplygin gas (MCG) in $f(R,T)$ gravity with particle creation. The
deterministic solutions to Einstein field equations are obtained here by
considering a simple parametrization of the Hubble parameter $H$ leading to
diverge behavior of scale factor. We have examine the physical behavior of
energy density, matter pressure, and the pressure due to particle creation
for the obtained models.

\begin{itemize}
\item In case of Model-I, the Universe exhibits an initial singularity of
the point-type at $t=0$. The model is well behaved in the range $0<t<\infty $
for the three cases of $\gamma $ = $0$, $0.5$ and $1$. It has been observed
that the Universe starts with infinite dominant energy density $\rho $ at
the initial singularity $t=0$, and monotonically decreases to attain a
finite value at the late time. The matter pressure $p_{m}$ decreases and
creation pressure $p^{c}$ increases as time increases and remains negative.
Thus, the model shows accelerating expansion of the Universe for $\beta >1$
and the rate of particle production are high initially which gradually and
stop at late times. The spatial volume increases exponentially with time
which indicates that the Universe starts its expansion with zero volume and
attains an infinite volume at late time.

\item In case of Model-II, the Universe exhibits point type initial
singularity $t=0$ and a future singularity of point-type at $t=\alpha $. The
model is well behaved in the range $0<t<\alpha $ for the three cases of $%
\gamma $ = $0$, $0.5$ and $1$. It has been observed that the Universe starts
with infinite dominant energy density $\rho $. As the time increases, the
energy density decreases and diverges towards negative in future
representing a future finite time singularity at $t=\alpha $. The matter
pressure $p_{m}$ decreases and after some time it falls down rapidly and
remains negative throughout the evolution of the Universe. Here, we observe
that the Universe is expanding with acceleration at late times and the rate
of particle production are initially high and later decreases to almost
negligible, and again increases to attain high rate of particle production.
The spatial volume increases with time which indicates that the Universe
starts its expansion with zero volume and attains a finite volume at late
time.

\item In case of Model-III, the Universe is bouncing in nature and is free
from initial singularity. It has been observed that the Universe starts with
finite energy density which increases rapidly to its maximum value then
gradually decreases to its smaller finite value. The matter pressure $p_{m}$
begins with a finite negative value remains negative forever. Here, this
model starts with a finite acceleration and the rate of acceleration
deceases with time. The rate of particle production are initially
negligible, and again increases to attain high rate of particle production,
and then gradually decreases at late time. The Universe starts its expansion
with finite volume at $t=0$ and attains an infinite volume at late time. The
model starts with infinite acceleration and decreases with time. The
deceleration parameter takes finite positive value at late time. Here, the
choice of the parameter $\beta $ is to be taken care for the positivity
condition of energy density $\rho $ and explaining the present Universe.

\item We have discussed the diverging behavior of different dark energy
models-I, II and III using statefinder pair $\{r,s\}$. The model-I is a
power law model in which $q$, $r$, $s$ are constants and depends on model
parameter $\beta $ only. The model describes $\Lambda CDM$ and $SCDM$ when $%
\beta \rightarrow $ $\infty $ and $\beta =\frac{2}{3}$ respectively. Thus,
the model-I describes for accelerating Universe when $\beta >1$ and we have
different expansion factors for different values of $\beta $ (see Table-1).

\item In the model-II, the parameters $q$, $r$, $s$ are time varying and
contains two model parameters $\alpha $ and $\beta $. To analyze behaviors
of our model better, we plot the trajectories of the models in $s-r$ and $%
q-r $ planes. Fig. 5a and Fig. 5b depict the evolution of trajectories with
time in $s-r$ plane and $q-r$ plane respectively. Fig. 4a depicts the model
converges from a fixed point $SCDM$ ($s=1$, $r=1$) to a fixed point $\Lambda
CDM$ ($s=0$, $r=1$). Fig. 4b depicts the model converges to the steady state
model (SS) ($q=-1$, $r=1$) and transform from the fixed point $SCDM$ ($q=%
\frac{1}{2}$, $r=1$) to $\Lambda CDM$ and end up with the fixed point $SS$.

\item In the model-III, Fig. 6a displays the evolution of trajectory with
time in the $s-r$ diagram. The statefinder ($s-r$) is shown for two
different values of model parameters $\alpha $ and $\beta $. The solid black
trajectory does not begins with $SCDM$ but passes through $\Lambda CDM$. The
dashed line trajectory starts with $SCDM$ converges to $\Lambda CDM$. Fig.
6b depicts that both trajectories evolve with the $SCDM$ but never touches
to the $SS$ model. Fig. 6a depicts that both trajectories do not start with $%
SCDM$ but both converge to $\Lambda CDM$ in $s-r$ plane. Fig. 6(b) depicts
that neither both trajectories start with $SCDM$ nor both converge to $SS$.

\item In Section 5, the stability condition $0\leq C_{s}^{2}\leq 1$ of the
all the models-I, II, III for the three cases of $\gamma $ = $0$, $0.5$ and $%
1$ has been examined. The plots of Fig. 7a and Fig. 7c represent the model-I
and III are perfectly stable for all $t$ but model-II is unstable due to
Big-Rip singularity at $t=\alpha $. Thus, we conclude that models-I and III
of the Universe completely stable and model-II is conditionally stable in $%
f(R,T)$ gravity with particle creation.

\item In Section 6, we have studied the lookback time, proper distance,
luminosity distance, angular diameter distance for our obtained models-I,
II, III through the plots in Fig. 8, 9. We can see, the model-III behaves
well in very small redshifts ($z<0.2$) while model-I and model-II show
better behavior for higher redshifts also ($z>>1$).

\item In subsection 6.5, we have discussed the phases of evolution of
deceleration parameter. The DP is constant throughout the evolution for
model-I while it is time varying for model-II and model-III. The model-III
can exhibit phase transition from acceleration to deceleration or eternal
acceleration while model-II can have deceleration to acceleration phase
transition explaining the current observation. We can see, for ($\alpha
=-6,\ \beta =-4$), the phase transition occurs at $z_{tr}=0$ and $q_{0}=0$
and the phase transition can be prepond by increasing $\alpha $ only i.e.
for ($\alpha =-5.55,\ \beta =-4$), we have $z_{tr}=0.29$ \& $q_{0}=-0.14$
and for ($\alpha =-5.1,\ \beta =-4$), we have $z_{tr}=0.72$ \& $q_{0}=-0.23$.
\end{itemize}

\vskip0.2in \noindent \textbf{Acknowledgements } The authors express their
thanks to CTP, Jamia Millia Islamia, New Delhi, India where a part of this
work have been done and also thankful to Prof. M. Sami for a fruitful
discussion to improve the paper. Author SKJP wishes to thank NBHM (DAE) for
financial support through the post doctoral research fellowship. 

\end{document}